\documentclass[12pt]{article}
\usepackage{amsmath}
\usepackage{graphicx}
\usepackage{enumerate}
\usepackage{natbib}
\usepackage{url} 

\usepackage{amsfonts}
\usepackage{amsthm}
\usepackage{multirow}   
\usepackage{setspace}
\usepackage{amssymb}
\usepackage{amsmath}
\usepackage{graphicx,color,epstopdf}
\usepackage{hyperref}
\usepackage{xcolor}
\usepackage{booktabs}
\usepackage{subfigure}
\usepackage{array}

\usepackage{dsfont }  
\usepackage{algorithm}  
\usepackage{algorithmic}
\usepackage{enumerate}     
\usepackage{multirow}
\usepackage{setspace}

\usepackage{float}

\usepackage{graphicx}

\newcommand{\blind}{1}
\newtheorem{theorem}{Theorem}[section]
\newtheorem{prop}{Proposition}[section]

\newtheorem{remark}{Remark}[section]
\newtheorem{definition}{Definition}[section]

\numberwithin{equation}{section}

\begin{document}

\def\spacingset#1{\renewcommand{\baselinestretch}%
{#1}\small\normalsize} \spacingset{1}


\if1\blind
{
	\title{\bf Corrected kernel principal component analysis for model structural change detection
}
\author{Luoyao Yu$^{1}$, Lixing Zhu$^{2}$, Ruoqing Zhu$^{3}$ and Xuehu Zhu$^{1}$
	\footnote{Corresponding author (X. Zhu). Email addresses: zhuxuehu@xjtu.edu.cn (X. Zhu).}\\
	$^1$ School of Mathematics and Statistics, Xi'an Jiaotong University, China\\
	$^2$ Center for Statistics and Data Science, Beijing Normal University, China\\
	$^3$ Department of Statistics, University of Illinois Urbana-Champaign, USA\\
}
  \maketitle
} \fi

\if0\blind
{
  \bigskip
  \bigskip
  \bigskip
  \begin{center}
    {\LARGE\bf Corrected kernel principal component analysis for model structural change detection}
\end{center}
  \medskip
} \fi

\bigskip
\begin{abstract}
		This paper develops a method to detect model structural changes by applying a Corrected Kernel Principal Component Analysis (CKPCA) to construct the so-called central distribution deviation subspaces. This approach can efficiently identify the mean and distribution changes in these dimension reduction subspaces. We derive that the locations and number changes in the dimension reduction data subspaces are identical to those in the original data spaces. Meanwhile, we also explain the necessity of using CKPCA as the classical KPCA  fails to identify the central distribution deviation subspaces in these problems. Additionally, we extend this approach to clustering by embedding the original data with nonlinear lower dimensional spaces, providing enhanced capabilities for clustering analysis. The numerical studies on synthetic and real data sets suggest that the dimension reduction versions of existing methods for change point detection and clustering significantly improve the performances of existing approaches in finite sample scenarios.
\end{abstract}

\noindent%
{\it Keywords:}  Change point detection; Central distribution deviation subspace; Clustering;  Dimension reduction functional subspace; Kernel principal component analysis.
\vfill

\newpage
\spacingset{1.9} 
\section{Introduction}
\label{sec:intro}

The recognition of alterations in data structures, which include changes in mean values and distributions, as well as clustering, constitutes a significant area of scholarly inquiry. This subject has elicited interest across a wide range of disciplines including economics, genetics, medicine, image analysis, network data, and public health, as evidenced in the work of \cite{Mihaela2010Modelling}, \cite{chen2012parametric}, \cite{cleynen2014comparing}, \cite{kirch2015detection}, \cite{2015Using}, and \cite{Gre2020A}.

Given the established methodologies for low-dimensional data (as thoroughly reviewed by \cite{niu2016multiple}), several nonparametric change-point detection methods have been proposed to identify distributional changes. \cite{zou2014} constructed a nonparametric maximum likelihood approach. Building upon Euclidean distances, \cite{matteson2014a} developed the E-Divisive method. The MultiRank method, focusing on rank, was put forth by \cite{multirank2015}. \cite{kcp2019} formulated a kernel multiple change-point (KCP) algorithm to identify change points, and \cite{Madrid2022} considered a unique algorithm grounded on kernel density estimation.

For high-dimensional data, numerous methods for high-dimensional change point detection in the literature are mainly for means. A considerable proportion of these methods employ the cumulative sum (CUSUM) statistics, as originally developed in \cite{cusum1997}. Aggregation across different dimensions of CUSUM statistics has proven to be a popular and effective strategy. \cite{jirak2015uniform} introduced the coordinate-wise CUSUM-statistics. The sparsified binary segmentation (SBS) method, proposed by \cite{cho2015multiple-change-point}, enables the detection of changes in high-dimensional time series. \cite{wang2018high} developed a projection-based method under the assumption of sparsity. \cite{enikeeva2019high} introduced a scan-statistic-based algorithm for detecting high-dimensional change points with sparse alternatives. In a different approach, \cite{wang2022inference} employed self-normalized U-statistics as an alternative to CUSUM statistics. But to the best of our knowledge, there are few methods available to detect distribution changes in high-dimensional scenarios.

The process of handling functional or high-dimensional data in dimension reduction subspaces is vital for mitigating the curse of dimensionality. Thus, a natural idea is to extend the methods handling low-dimensional data to high-dimensional scenarios in low-dimensional subspaces. Principal Component Analysis (PCA) is a popularly used method. For instance, \cite{kuncheva2012pca} examined changes in streaming data via PCA, focusing on the mean and covariance matrix. \cite{qahtan2015pca} fused a semi-parametric log-likelihood change detector with PCA when dealing with streaming data. Additionally, \cite{jiao2021change} crafted a spectral PCA change point method specifically for multivariate time series. Despite these advancements, these studies have not provided theoretical evaluations to explicate why PCA is an effective strategy for this problem, nor have they investigated whether dimension reduction subspaces may result in the loss of information pertaining to the change structure of the original data.

In response to these gaps in current understanding, this paper takes into account distributional changes, regarding alterations in means as a particular case. To facilitate this, we propose a method called Corrected Kernel Principal Component Analysis (CKPCA), designed to identify dimension reduction subspaces - herein referred to as central distribution deviation subspaces. This enables the identification of distributional changes within these subspaces. Furthermore, it is demonstrated that both the locations and the number of changes are consistent between original data spaces and CKPCA. As a corollary, the detection of changes in the structure for means is treated as a specific case.

By carrying out further detection within the lower-dimensional subspace, a substantial improvement can be made to the performance of existing methods dealing with high-dimensional data. This is explored in numerical studies where we evaluate the performance of various pre-existing methodologies as previously discussed. Moreover, we implement an iterative subspace clustering algorithm built upon CKPCA. This serves to enhance classic clustering techniques like the K-means method \citep{kmeans}, the expectation-maximization algorithm \citep{EM}, and the density-based spatial clustering of applications with noise method \citep{dbscan}.

The structure of the remaining sections of this paper is as follows: Section 2 introduces the concept of the central distribution deviation subspace, while proposing the use of the CKPCA method for its identification. Section 3 includes the application of our dimension reduction technique in clustering and presents an iterative algorithm. Section 4 features simulation studies and the analysis of several real data sets to illustrate our findings. Section 5 analyzes the benefits and drawbacks of the new method, along with additional research areas. Supplementary Materials discuss a special case where $\phi(X_t)=X_t$, reducing CKPCA to CPCA. It also includes simulations with mean changes, analysis of Macroeconomic data, regularity conditions, and technical proofs for the theorems.

\section{Corrected kernel principal component analysis}

Given a set of $p$-dimensional observations $X_i = (X_{i1}, \ldots, X_{ip})^\top \in \mathbb{R}^p$, where $\mu_i = E(X_i)$ and $\Sigma_i = \text{Cov}(X_i)$ for $i = 1, 2, \ldots, n$. We assume that $\{X_i\}_{i=1}^n$ follows unknown distributions $\{P_i\}_{i=1}^n$, where $P_i$ represents an unknown probability distribution without any parametric prior. Our objective is to identify $s$ change points $1 \leq z_1 < z_2 < \ldots < z_s \leq n$ that satisfy the condition:
\begin{equation}
	P_{z_{i}+1}=\ldots=P_{z_{i+1}}=: P^{(i)} \ {\rm{and}}\ P^{(i)}\neq P^{(i+1)}, \quad \forall 0 \leqslant i \leqslant s.
	\label{change2}
\end{equation}
The model (\ref{change2}) represents changes in the distribution. Similar to \cite{2017New} and \cite{kcp2019}, we employ a nonlinear feature map $\phi$ to transform $X_t$ as follows:
$$
X_t \rightarrow \phi(X_t)=Y_t, t=1,2,...,n,
$$
where $\phi: \mathcal{X} \rightarrow \mathcal{H}$ and $\mathcal{H}$ is a reproducing kernel Hilbert space (RKHS). We consider the model proposed by \cite{kcp2019}:
\begin{equation}\label{model}
	\forall 1 \leq t \leq n, \quad Y_{t}=\Phi\left(X_{t}\right)=\mu_{t}^{\star}+\epsilon_{t} \in \mathcal{H},
\end{equation}
where $\mu_{t}^{\star}$ is the ``mean'' element of $\Phi(X_t)$ and $\epsilon_{t}=Y_t-\mu_{t}^{\star}$. According to \cite{ledoux1991probability}, if $\mathcal{X}$ is separable and $\mathbb{E}\left[K\left(X_{t}, X_{t}\right)\right]<+\infty$, then $\mu_{t}^{\star}$ exists and is uniquely defined as an element in $\mathcal{H}$:
\begin{equation}
	\forall f \in \mathcal{H}, \quad\left\langle\mu_{t}^{\star}, f\right\rangle_{\mathcal{H}}=\mathbb{E}\left\langle\Phi\left(X_{t}\right), f\right\rangle_{\mathcal{H}},
\end{equation}
where $\langle\cdot, \cdot\rangle_{\mathcal{H}}$ represents the inner product in $\mathcal{H}$.

Furthermore, when $K$ is a characteristic kernel such as the Gaussian kernel \citep{fukumizu2004dimensionality}, any change in the distribution $P_{z_i}$ implies a change in the mean element $\mu_{z_{i}}^{\star}$. Therefore, the distributional change in the model (\ref{change2}) can be transformed into a mean change problem:
$$
\mu_{z_{i}+1}^{\star}=\ldots=\mu_{z_{i+1}}^{\star}=: \mu_{d}^{(i)}  \ {\rm{and}}\ \mu_{d}^{(i)} \neq \mu_{d}^{(i+1)}  , \quad \forall\ 0 \leqslant i \leqslant s.
$$	
Based on this result, we introduce the concept of the central distribution deviation subspace.
\begin{definition}
	${\rm{Span}}\{\mu_{d}^{(i)}-\mu_{d}^{(j)}, \ {\rm{for}} \ i, j= 1, \cdots,s+1 \}$ is called the central distribution deviation subspace of the sequence $\{X_i\}_{i=1}^n$ and  is written as $\emph{S}^{d}_{\{X_i\}_{i=1}^n}$. For this functional subspace, $q_d={\rm{dim}}\{ \emph{S}^{d}_{\{X_i\}_{i=1}^n} \}$ is called the structural dimension of  $\emph{S}^{d}_{\{X_i\}_{i=1}^n}$.
\end{definition}

	It is worth noting that the structural dimension $q_d$ is unknown and satisfies $q_d\leq s$. The following theorem guarantees the integrity of information of the original high-dimensional data in the lower-dimensional subspace.	
\begin{theorem}
	For any basis functions $\{v_1,v_2,...,v_{q_d}\}$ of $\emph{S}^{d}_{\{X_i\}_{i=1}^n}$ with $q_d\le s$, let $f(X_i)=\left(\left\langle v_1, Y_i\right\rangle_{\mathcal{H}}, \cdots,\left\langle v_{q_d}, Y_i\right\rangle_{\mathcal{H}}\right)$.
	Both the sequences $\{ f(X_i) \}_{i=1}^n$ and $\{X_i\}_{i=1}^n$ have the same locations of changes.
\end{theorem}

Kernel Principal Component Analysis (KPCA) is an efficient method for nonlinear dimension reduction, serving as the nonlinear counterpart to PCA. Section 12 of \cite{li2018} generalized ``the sample covariance matrix'' to ``the sample covariance operator,'' denoted as in our case:
\begin{equation}
	\begin{aligned}
		M_{n}^{kernel}
		& =\frac{1}{n} \sum_{i=1}^{n}\left(Y_i-\bar{Y}\right) \otimes   \left(Y_i-\bar{Y}\right),
	\end{aligned}
\end{equation}
where the tensor product $f \otimes g$ is the operator on $\mathcal{H}$ such that $(f \otimes g) h=f\langle g, h\rangle_{\mathcal{H}}$ for all $h \in \mathcal{H}$ and two members $f$ and $g$ of $\mathcal{H}$. However, working directly on $M_n^{kernel}$ is insufficient for identifying the true locations and the number of changes, as this operator fails to account for change-related information. By computing the expectation of $M_n^{kernel}$, we observe that
\begin{equation}\label{t1}
	\begin{aligned}
		E(M_n^{kernel})&=\sum_{j=1}^{s}\frac{1}{n} \sum_{i=z_{j}+1}^{z_{j+1}} E\left\{\left(Y_{i}-\bar{Y}\right) \otimes \left(Y_{i}-\bar{Y}\right)\right\} \\
		&=\sum_{j=1}^{s}\frac{1}{n} \sum_{i=z_{j+1}}^{z_{j+1}} E\Bigg\{\left(Y_{i}-\sum_{l=1}^{s}c_l \mu_{d}^{(l)}-\sum_{l=1}^{s} c_{l}\left(\bar{Y}_{l}-\mu_{d}^{(l)}\right)\right)\\
		&\ \ \ \ \ \ \ \ \ \ \ \ \otimes \left(Y_{i}-\sum_{l=1}^{s}c_l \mu_{d}^{(l)}-\sum_{l=1}^{s} c_{l}\left(\bar{Y}_{l}-\mu_{d}^{(l)}\right)\right)\Bigg\} \\
		&=\sum_{j=1}^{s} c_j \Sigma^{(j)}_d+\sum_{j=1}^{s} \sum_{i=1}^{s} \sum_{m=1}^{s} c_jc_ic_m \left(\mu_{d}^{(j)}-\mu_{d}^{(i)}\right)\otimes
		\left(\mu_{d}^{(j)}-\mu_{d}^{(m)}\right)+\frac{s}{n}\sum_{l=1}^{s}c_l \Sigma^{(l)}_d \\
		&\rightarrow \sum_{j=1}^{s} c_j \Sigma^{(j)}_d+\sum_{i=1}^{s} \sum_{j=1}^{s} c_ic_j\left(\mu_{d}^{(i)}-\mu_{d}^{(j)}\right)\otimes \left(\mu_{d}^{(i)}-\mu_{d}^{(j)}\right)\\
		&=:\Sigma_{pooled}^{kernel}+\Delta^{kernel}=M^{kernel},
	\end{aligned}
\end{equation}
where $n_i /n\rightarrow c_i$ as $n\rightarrow \infty$, and $n_i$ is the length of the segment between two consecutive changes. $\Sigma_d^{(i)}$ refers to the covariance operator of $Y_j$, where $j=z_{i}+1,\cdots,z_{i+1}$. Moreover, $\Sigma_{pooled}^{kernel}=\sum_{j=1}^{s} c_j \Sigma^{(j)}_d$ and $\Delta^{kernel}=\sum_{i=1}^{s}\sum_{j=1}^{s} c_ic_j(\mu_{d}^{(i)}-\mu_{d}^{(j)})\otimes (\mu_{d}^{(i)}-\mu_{d}^{(j)})$.
This demonstrates that the corrected operator $\Delta^{kernel}$ can fully provide information about the changes. The following theorem states that the eigenfunctions of $\Delta^{kernel}$ span the central distribution deviation subspace. Let $\operatorname{ran}(\Delta^{kernel})$ represent the range of $\Delta^{kernel}$, and $\overline{\operatorname{ran}}(\Delta^{kernel})$ be its closure. We refer to KPCA based on $\Delta^{kernel}$ as corrected KPCA, and investigate the relationship between $\overline{\operatorname{ran}} (\Delta^{kernel})$ and $\emph{S}^{d}_{\{X_i\}_{i=1}^n}$.

\begin{theorem}\label{deltaZ}
	Under the model (\ref{model}), $\overline{\operatorname{ran}} (\Delta^{kernel})= \emph{S}^{d}_{\{X_i\}_{i=1}^n}$. Furthermore, letting $v_1,\cdots,v_{q_{d}}$ denote the eigenfuctions of $\Delta^{kernel}$ associated with the nonzero eigenvalues of $\Delta^{kernel}$,   ${\rm{Span}}(v_1,v_2,\cdots,v_{q_d}) = \emph{S}^{d}_{\{X_i\}_{i=1}^n}$.
\end{theorem}

Theorem \ref{deltaZ} emphasizes the importance of having a reliable estimator for the pooled covariance matrix $\Sigma^{kernel}_{pooled}$ to identify the subspace $\emph{S}^{d}_{\{X_i\}_{i=1}^n}$. To achieve this, we employ a localized approach to estimate $\Sigma^{kernel}_{pooled}$ as follows. Divide the data into $r$ segments: $\mathcal{S}_m=\{{(m-1)\alpha_n+1},\cdots,{m\alpha_n}\}$ for $m=1,2,\ldots,r-1$, and $\mathcal{S}_{r}=\{{(r-1)\alpha_n+1},\cdots,{n}\}$, where $ r=\lfloor n/\alpha_n \rfloor$. Compute the covariance matrices for each segment and then average them to obtain the final estimator $\Sigma^{kernel}_{pooled,n}$ of $\Sigma^{kernel}_{pooled}$:
\begin{eqnarray}
	\Sigma_{pooled,n}^{kernel}={{1\over r}\sum_{m=1}^{r} \hat{\Sigma}_m}\ {\rm{with}}\
	\hat{\Sigma}_m=\frac{1}{\hat{n}_m-1} \sum_{i\in \mathcal{S}_m } (Y_i-\bar{Y}_m)\otimes (Y_i-\bar{Y}_m),
\end{eqnarray}
where $\bar{Y}_m=\frac{1}{\hat{n}_m} \sum_{k\in \mathcal{S}_m }  Y_{k}$ with $\hat{n}_m$ being the cardinality of the sets $\mathcal{S}_m$'s.
$\Delta^{kernel}$ can be estimated as:
\begin{eqnarray}\label{3.5}
	\Delta_n^{kernel}=M_n^{kernel}-\Sigma_{pooled,n}^{kernel}.
\end{eqnarray}

\begin{theorem}\label{kernel}
	If $E\{K(X_i,X_i)\}$ and $K(X_i,X_i)$ are bounded, for $i=1,\cdots, n$, then $\Delta^{kernel}$ is a Hilbert-Schmidt operator and
	$$ \vert\vert\Delta_n^{kernel}-\Delta^{kernel}\vert\vert_{HS}=O_p\left(\sqrt{\frac{1}{n}}+\frac{\alpha_n}{n} \right),
	$$
	where $\vert\vert\cdot\vert\vert_{HS}$ denotes the Hilbert-Schmidt norm. Furthermore, let $\hat{P}_k$ and $P_k$ be the projection operators onto the subspaces spanned by the $k$th eigenfunctions of $\Delta_n^{kernel}$ and $\Delta^{kernel}$, respectively, for $k=1, \ldots, q_{d}$. Then
	$$
	\vert\vert \hat{P}_k-P_k \vert\vert_{\mathcal{H}} =O_p\left(\sqrt{\frac{1}{n}}+\frac{\alpha_n}{n} \right).
	$$
\end{theorem}

\begin{remark}
	From the formula (\ref{t1}), KPCA is equivalent to learning the subspace $\emph{S}^{d}_{\{X_i\}_{i=1}^n}$ when $\Sigma^{kernel}_{pooled}=\sigma I$, where $I$ is the identity operator and $\sigma$ is a constant. However, if $\Sigma^{kernel}_{pooled}\neq \sigma I$ for any $\sigma$, the dimension reduction subspace of KPCA involves both $\Sigma^{kernel}_{pooled}$ and $\Delta^{kernel}$. As a result, the original data sequence's change structures may not be preserved in the low-dimensional data sequence obtained through KPCA. This invalidates the use of KPCA for change point detection. The CKPCA algorithm addresses this issue by ensuring the preservation of the change structure's integrity during dimension reduction.
\end{remark}

\begin{remark}
	Theorem \ref{kernel} states that when $\alpha_n=o\left(n^{m}\right)$, with $0 \leq m \leq 1/2$ (including the case where the sample size $\alpha_n$ is finite), we can ensure
	$$
	\vert \vert 	\hat{P}_k-P_k\vert\vert_{\mathcal{H}}=O_p\left(\sqrt{\frac{1}{n}}\right).
	$$
	However, there is no theoretical criterion for selecting an optimal parameter $\alpha_n$ in practical scenarios. If $\alpha_n$ is too small, the estimated covariance for each segment may not be accurate enough, resulting in a lossy estimator of the pooled covariance matrix $\Sigma^{kernel}_{pooled}$. Conversely, if $\alpha_n$ is too large and a long segment may contain multiple distributions, the estimator may fail to identify the changes accurately. To strike a balance, numerical studies in the later section support choosing $\alpha_n= \lfloor \sqrt{n} \rfloor $.	 Note that the estimator is asymptotically unbiased, and thus, selecting $\alpha_n$ is intrinsically different from selecting a bandwidth in nonparametric estimation that can have an optimal selection when balancing between the bias and variance.
\end{remark}

The	formula (\ref{3.5}) requires solving the eigenvalue problem in the feature space ${\rm{Span}}\{v_1,\cdots,v_{q_d}\}$ to obtain the appropriate lower-dimensional data $f(X)$. However, since we lack knowledge about the explicit solution of $Y$, we employ a kernel trick to compute the eigenfunctions of $\Delta_n^{kernel}$. In this approach, we introduce the kernel matrix $K$, where $K_{ij}=\langle Y_i,Y_j\rangle_{\mathcal{H}}=K(X_i,X_j)$. Here, $K(\cdot,\cdot)$ represents a kernel function. Several kernel functions, including Gaussian, Laplace, and exponential kernels, satisfy the assumptions required for the kernel function in Theorem \ref{kernel}.

Then, $\Delta_n^{kernel}$ can be  expressed as
\begin{equation*}
	\begin{aligned}
		M_n^{kernel}-\Sigma_{pooled,n}^{kernel}
		&=\frac{1}{n}\sum_{i=1}^{n} (Y_i-\bar{Y})\otimes(Y_i-\bar{Y})- {{1\over r}\sum_{m=1}^{r} \hat{\Sigma}_m}\\
		&=\frac{1}{n}\sum_{i=1}^{n} (Y_i-\bar{Y})\otimes(Y_i-\bar{Y})-{1\over r} \sum_{m=1}^{r}
		\frac{1}{\hat{n}_m-1} \sum_{i\in \mathcal{S}_m } (Y_i-\bar{Y}_m)\otimes(Y_i-\bar{Y}_m).
	\end{aligned}
\end{equation*}
Let $\hat{v}_i$ and $\lambda_i$ represent the eigenfunction and eigenvalue of $\Delta_n^{kernel}$, respectively. For any $\hat{v}_i$ and $\lambda_i \neq 0$, we have the relationship where $\hat{v}_i$ and $Y$ satisfy:
\begin{equation}
	\begin{aligned}
		\lambda_i \hat{v}_i&=\Delta_n^{kernel}\hat{v}_i \\
		\Rightarrow   \lambda_i \hat{v}_i&=
		\frac{1}{n}\sum_{i=1}^{n} (Y_i-\bar{Y}) \langle Y_i-\bar{Y}, \hat{v}_i\rangle_{\mathcal{H}}-{1\over r} \sum_{m=1}^{r}
		\frac{1}{\hat{n}_m-1} \sum_{i\in \mathcal{S}_m } (Y_i-\bar{Y}_m)
		\langle Y_i-\bar{Y}_m, \hat{v}_i\rangle_{\mathcal{H}}.
	\end{aligned}
	\label{k}
\end{equation}
This suggests that $\hat{v}_i$ is a linear expression of $Y$. Therefore, we can set $\hat{v}_i=Y \alpha_i$. Define
$G_i=\left[\begin{array}{c}
	0_{i \alpha_n \times \alpha_n} \\
	I_{\alpha_{n} \times \alpha_{n}} \\
	0_{(n-(i+1)\alpha_n) \times \alpha_n}
\end{array}\right]$ and $H_i=\left[\begin{array}{c}
	0_{i \alpha_n \times \alpha_n} \\
	1_{\alpha_{n} \times \alpha_{n}} \\
	0_{(n-(i+1)\alpha_n) \times \alpha_n}
\end{array}\right]$,
where $I_{p \times p}$ represents the identity matrix, while $1_{p \times p}$ denotes the $p$-dimensional matrix with all elements equal to 1. We can then solve the eigen-decomposition problem by substituting the kernel matrix into the following inference:
\begin{equation}
	\begin{aligned}
		&\Delta_n^{kernel}\hat{v}_i=\lambda_i \hat{v}_i \\
		\Rightarrow &\left[\frac{1}{n}\sum_{i=1}^{n} (Y_i-\bar{Y})\bar{K}_i-{1\over r} \sum_{m=1}^{r}
		\frac{1}{\hat{n}_m-1} \sum_{i\in \mathcal{S}_m } (Y_i-\bar{Y}_m)\bar{K}_i^{(m)}  \right] \alpha_i=\lambda_i Y \alpha_i \\
		\Rightarrow &\left[\frac{1}{n}\sum_{i=1}^{n} \bar{K}_i^{T}\bar{K}_i-{1\over r} \sum_{m=1}^{r}
		\frac{1}{\hat{n}_m-1} \sum_{i\in \mathcal{S}_m } (\bar{K}_i^{(m)})^{T}\bar{K}_i^{(m)} \right] \alpha_i=\lambda_i K \alpha_i \\
		\Rightarrow & (KLK-KUK) \alpha_i=\lambda_i K \alpha_i \\		
		\Rightarrow & K_n \alpha_i=\lambda_i \alpha_i,
	\end{aligned}
	\label{correctk}
\end{equation}
where
\begin{equation*}
	\begin{aligned} \bar{K}_i=&K_{i1}-\frac{1}{n}\sum_{j=1}^{n}K_{j1},\cdots,K_{in}-\frac{1}{n}\sum_{j=1}^{n}K_{jn},\\ \bar{K}_i^{(m)}=&K_{i1}-\frac{1}{\hat{n}_m}\sum_{j\in \mathcal{S}_m}K_{j1},\cdots,K_{in}-\frac{1}{\hat{n}_m}\sum_{j\in \mathcal{S}_m}K_{jn},\\
	L=&\frac{1}{n} (I_{n \times n}- \frac{1}{n}1_{n \times n})(I_{n \times n}- \frac{1}{n}1_{n \times n})^{\top}, \\
	U=& \sum_{i=1}^{r} \frac{1}{r (n_i-1)} (G_i-\frac{1}{\alpha_{n}}H_i)(G_i-\frac{1}{\alpha_{n}}H_i)^{\top},\\
	K_n=&(L-U)K.
	\end{aligned}
\end{equation*}
Based on the formula (\ref{correctk}), we have that $\alpha_i$ is an eigenvector of $K_n$. For any $i=1,2,\cdots,n$, we have:
\begin{equation}
	\begin{aligned}
		f(X_i)&=\left(\left\langle v_1, Y_i\right\rangle_{\mathcal{H}}, \cdots,\left\langle v_{q_d}, Y_i\right\rangle_{\mathcal{H}}\right) \\
		&=\left(\left\langle Y\alpha_1, Y_i\right\rangle_{\mathcal{H}}, \cdots,\left\langle Y\alpha_{q_d}, Y_i\right\rangle_{\mathcal{H}}\right)\\
		&=(K_i\alpha_1,\cdots,K_i\alpha_{\hat{q}_d}) \\
		&=K_i B_n,
	\end{aligned}
\end{equation}
where $K_i=(K_{i1},K_{i2},\cdots,K_{in})$ and $B_n=(\alpha_1,\alpha_2,\dots,\alpha_{q_d})$. Therefore, the data after dimension reduction is given by $f(X)=(f_1(X)^{T},\cdots,f_n(X)^{T})=KB_n$. It can be observed that $\Delta_n^{kernel}$ and $K_n$ share the same non-zero eigenvalues.
To estimate the structural dimension $q_d$ when it is unknown, we employ the thresholding ridge ratio criterion (TRR) as follows:
\begin{eqnarray}\label{hatq}
	\hat{q}_d:=\max_{1\leq k \leq n-1}\left\{k: \ \hat r_k=\frac{\hat{\lambda}_{k+1}+c_n}{\hat{\lambda}_k+c_n} \leq \tau \right\},
\end{eqnarray}
where $\hat{\lambda}_{1} \geq \ldots \geq \hat{\lambda}_{n}$ are the eigenvalues of the estimated target matrix $K_n$, $c_n$ is a ridge value approaching zero at a certain rate, and $\tau$ is a thresholding value such that $0<\tau <1$. Choosing $\tau=0.5$ is reasonable according to the plug-in principle in \cite{zhu2020Dimensionality} to avoid overestimation with a large $\tau$ and underestimation with a small $\tau$. As the target matrix differs from \cite{zhu2020Dimensionality}, there is no optimal criterion or theoretical result for selection. However, we recommend setting the ridge value to $c_n=0.2\log (\log(n)) \sqrt{1/n}$ for practical purposes. The consistency of $\hat{q}_d$ is stated in the following theorem.
\begin{theorem}\label{q-identify}
	Let $\tilde{\eta}_n=\max\left\{\sqrt{\frac{1}{n}},\frac{\alpha_n}{n}\right\}.$
	Under the same conditions in Theorem~\ref{kernel}, if $c_n$ satisfies $c_n\to 0$, $\tilde{\eta}_n \to 0$, $c_n/\tilde{\eta}_n\to \infty$ as $n\to \infty$, then $P\left(\hat{q}_d=q_d\right)\to 1$.
\end{theorem}

In the specific scenario where $\phi(X_t)=X_t$, KPCA simplifies to PCA. Furthermore, we introduce Corrected PCA (CPCA) as a method to address mean changes. For the sake of brevity, detailed information about CPCA is presented in the Supplementary Materials.

\section{Iterative CKPCA in cluster analysis}

Clustering algorithms frequently involve the calculation of distances between observations. However, the accuracy of these distance calculations can be influenced by factors such as dimensionality and nonlinearity. To address this, dimension reduction is commonly employed in clustering analysis, serving both visualization and accuracy purposes (UMAP \citep{mcinnes2018umap}, t-SNE \citep{van2008visualizing}). Among the popular dimension reduction approaches in cluster analysis are PCA and KPCA. However, determining the optimal number of directions can be challenging and can significantly impact the clustering results. In this section, we extend CKPCA to cluster analysis, aiming to achieve a superior nonlinear low-dimensional embedding. Although the approach is applicable to any clustering algorithm, we illustrate its effectiveness using the commonly used clustering methods such as K-means algorithm.

Consider a dataset $\{X_i\}_{i=1}^n$ where $X_i=(X_{i1},\cdots, X_{ip})^{\top} \in \mathbb{R}^{p}$ are independent. The data points belong to a union of $d$ categories denoted by $\{\mathcal{C}_k\}_{k=1}^{d}$, with each category $\mathcal{C}_k$ containing $n_k$ points such that $\sum_{k=1}^dn_k=n$. Assume $n_j/n\rightarrow \omega_j$ as $n\rightarrow \infty$ be the weight of category $\mathcal{C}_j$. We apply the nonlinear feature map $X_i\rightarrow \phi(X_i)=Y_i$, $i=1,\cdots,n$. There exists a $\phi$ such that if $X_i \in \mathcal{C}_k$ and $X_j\in \mathcal{C}_k$ for $k=1,\cdots, d$, then $E(Y_i)=E(Y_j)$ holds. For all $Y_j\in \mathcal{C}_k$, let $E(Y_j)=\mu^{(k)}_{d}$ and $\Sigma^{(k)}_{d}= E(Y_j-\mu^{(k)}_{d})\otimes (Y_j-\mu^{(k)}_{d}),$ for $k=1, \cdots, d$.

Consistent with the inference in Section 3,  define the central distribution deviation subspace in cluster analysis as $\emph{S}^{d}_{\{X_i\}_{i=1}^n}={\rm{Span}}\{\mu_{d}^{(i)}-\mu_{d}^{(j)}, \ {\rm{for}} \ i, j= 1, \cdots,d+1 \}$ with the structural dimension $q_d={\rm{dim}}\{ \emph{S}^{d}_{\{X_i\}_{i=1}^n} \}$. Consider the ``sample covariance operator'' $M^{kernel}_n$, and its expectation is
\begin{equation}\label{clu}
	\begin{aligned}
		E(M^{kernel}_n)&=\sum_{j=1}^{d}\frac{1}{n} \sum_{i \in \mathcal{C}_j} E\left\{\left(Y_{i}-\bar{Y}\right)\otimes
		\left(Y_{i}-\bar{Y}\right)\right\} \\	
		&=\sum_{j=1}^{d}\frac{1}{n} \sum_{i \in \mathcal{C}_j} E\Bigg\{\left(Y_{i}-\sum_{l=1}^{d}\omega_l \mu_{d}^{(l)}-\sum_{l=1}^{d} \omega_{l}\left(\bar{Y}_{l}-\mu_{d}^{(l)}\right)\right) \\
		&\ \ \ \ \ \ \ \ \otimes \left(Y_{i}-\sum_{l=1}^{d}\omega_l \mu_{d}^{(l)}-\sum_{l=1}^{d} \omega_{l}\left(\bar{Y}_{l}-\mu_{d}^{(l)}\right)\right)\Bigg\} \\
		&=\sum_{j=1}^{d} \omega_j \Sigma^{(j)}_{d}+\sum_{j=1}^{d} \sum_{i=1}^{d} \sum_{m=1}^{d} \omega_j\omega_i\omega_m (\mu_{d}^{(j)}-\mu_{d}^{(i)})\otimes(\mu_{d}^{(j)}-\mu_{d}^{(m)})+\frac{d}{n}\sum_{l=1}^{d}\omega_l \Sigma^{(l)}_{d} \\
		&\rightarrow \sum_{j=1}^{d} \omega_j \Sigma^{(j)}_{d}+\sum_{i=1}^{d} \sum_{j=1}^{d} \omega_i\omega_j(\mu_{d}^{(i)}-\mu_{d}^{(j)})\otimes(\mu_{d}^{(i)}-\mu_{d}^{(j)})\\
		&=:\Sigma_{pooled}^{kernel}+\Delta^{kernel}=M^{kernel},
	\end{aligned}
\end{equation}
where $\Delta^{kernel}=\sum_{i=1}^{d} \sum_{j=1}^{d} \omega_i\omega_j(\mu_{d}^{(i)}-\mu_{d}^{(j)})\otimes(\mu_{d}^{(i)}-\mu_{d}^{(j)})$.

\begin{prop}\label{clus1} 	For any basis functions $\{v_1,v_2,\cdots,v_{q_d}\}$ of $\emph{S}^{d}_{\{X_i\}_{i=1}^n}$ with $q_d\leq d$
	Let $f(X_i)=(\left\langle v_1, Y_i\right\rangle_{\mathcal{H}}, \cdots,\left\langle v_{q_d}, Y_i\right\rangle_{\mathcal{H}})$. 	
	Both the sequences $\{ f(X_i) \}_{i=1}^n$ and $\{X_i\}_{i=1}^n$ have the same clustering results.
\end{prop}

To estimate $\Sigma^{kernel}_{pooled}$, we use a weighted method given by:
\begin{eqnarray}\label{cluster-sigma}
	\Sigma^{kernel}_{pooled,n}={\sum_{i=1}^{d} \frac{n_i-1}{n-d} \hat{\Sigma}_{i}}\ {\rm{with}}\
	\hat{\Sigma}_{i}=\frac{1}{n_i-1} \sum_{j\in \mathcal{C}_i} (Y_j-\bar{Y}_i)\otimes(Y_j-\bar{Y}_i),
\end{eqnarray}
where $\bar{Y}_i=\frac{1}{n_i}\sum_{j \in \mathcal{C}_i} Y_j$. Then, we estimate $\Delta_n^{kernel}$ as follows:
\begin{equation}
	\begin{aligned}
		\Delta^{kernel}_n&=M^{kernel}_n-\Sigma^{kernel}_{pooled,n}\\
		&=\frac{1}{n}\sum_{i=1}^{n}(Y_i-\bar{Y})\otimes(Y_i-\bar{Y})- \sum_{i=1}^{d} \frac{n_i-1}{n-d} \hat{\Sigma}_{i}.
	\end{aligned}
	\label{cluster}
\end{equation}
Since the categories $\mathcal{C}_j$ are unknown, we first obtain an initial value by applying a popular clustering method after kernel PCA. We then propose an iterative approach, which will be described in detail later.

Given the categories $\mathcal{C}_j$, for each $X_j\in \mathcal{C}_i$, we define $G_i$ as an $n \times n_k$ matrix with a 1 at the $(\mathcal{C}_i(j),j)$th element and zeros elsewhere. Similarly, we define $H_i$ as an $n \times n_k$ matrix with a row of 1s at the $\mathcal{C}_i(j)$th position and zeros elsewhere, where $j=1,\cdots,n_i$ and $k=1,2,\cdots,d$. Following the approach in Section 3, let $\hat{v}_i$ denote an eigenfunction of $\Delta_n^{kernel}$, and we can express $\hat{v}_i$ as $\hat{v}_i=Y \alpha_i$ for some $\alpha_i$. For any $\hat{v}_i$ and $\lambda_i \neq 0$, we have:
\begin{equation}
	\begin{aligned}
		&\Delta_n^{kernel}\hat{v}_i=\lambda_i \hat{v}_i \\
		\Rightarrow & \left [ \frac{1}{n}\sum_{i=1}^{n}(Y_i-\bar{Y})\otimes(Y_i-\bar{Y})- \sum_{i=1}^{d} \frac{n_i-1}{n-d} \hat{\Sigma}_{i} \right ]
		Y\alpha_i=\lambda_i Y \alpha_i \\
		\Rightarrow  & \left[\frac{1}{n}\sum_{i=1}^{n} (Y_i-\bar{Y})\bar{K}_i-{1\over r} \sum_{m=1}^{r}
		\frac{1}{\hat{n}_m-d} \sum_{i\in \mathcal{C}_m } (Y_i-\bar{Y}_m)\bar{K}_i^{(m)}
		\right] \alpha_i=\lambda_i Y \alpha_i \\
		\Rightarrow &\left[ 	\frac{1}{n}\sum_{i=1}^{n} \bar{K}_i^{T}\bar{K}_i-{1\over r} \sum_{m=1}^{r}
		\frac{1}{\hat{n}_m-1} \sum_{i\in \mathcal{C}_m } (\bar{K}_i^{(m)})^{T}\bar{K}_i^{(m)}
		\right]\alpha_i
		=\lambda_i K \alpha_i \\
		\Rightarrow &(KRK-KSK) \alpha_i=\lambda_i K \alpha_i \\		
		\Rightarrow & K_n \alpha_i=\lambda_i \alpha_i,
	\end{aligned}
	\label{kernelclu}
\end{equation}
where $\bar{K}_i=(K_{i1}-\frac{1}{n}\sum_{j=1}^{n}K_{j1},\cdots,K_{in}-\frac{1}{n}\sum_{j=1}^{n}K_{jn})$, $\bar{K}_i^{(m)}=(K_{i1}-\frac{1}{\hat{n}_m}\sum_{j\in \mathcal{C}_m}K_{j1},\cdots,K_{in}-\frac{1}{\hat{n}_m}\sum_{j\in \mathcal{C}_m}K_{jn})$, $R=\frac{1}{n} (I_{n \times n}- \frac{1}{n}1_{n \times n})(I_{n \times n}- \frac{1}{n}1_{n \times n})^{\top}$, $S=\sum_{i=1}^{d} \frac{1}{n-d}  (G_i-\frac{1}{n_{i}}H_i)(G_i-\frac{1}{n_{i}}H_i)^\top$, $K_n=(R-S)K$ and $\alpha_i$ is the eigenvector of $K_n$.
Moreover, the lowered dimensional data is $f(X)=(f(X_1)^{T},\cdots,f(X_n)^{T})=KB_n$ and $D_{n}=(\alpha_{1},\cdots,\alpha_{\hat{q}_{d}})$.  In order to apply the formulas (\ref{cluster-sigma}) and (\ref{cluster}), it is important to have information about the categories $\{\mathcal{C}_k\}_{k=1}^{d}$.

However, since information about the categories $\mathcal{C}_j$ is lacking, we need to obtain an initial value by applying a popular clustering method after KPCA. Therefore, we propose an iterative algorithm as follows. According to Proposition \ref{clus1}, we set $\hat{q}_{d0}=d-1$ as the initial value. Then, we obtain the initial lowered dimensional data $f^{(0)}(X)=K B_{0n}$, where $B_{0n}=(\alpha^{0}_{1},\cdots,\alpha^{0}_{\hat{q}_{d0}})$ represents the eigenvectors associated with the largest $\hat{q}_k$ eigenvalues of the kernel matrix $K_{0n}=RK$. Note that $f^{(0)}(X)$ is equivalent to the lowered dimensional data in KPCA. Next, we use a popular clustering method based on $f^{(0)}(X)$ to obtain the initial categories $\{\hat{\mathcal{C}}_i\}_{i=1}^d$, with the pre-defined number of categories $d$. Finally, we apply the formulas (\ref{cluster}) and (\ref{kernelclu}) using the results $\{\hat{\mathcal{C}}_i\}_{i=1}^d$ to obtain the lowered dimensional data $f(X)=KB_n$, where $B_n=(\alpha_1,\alpha_2,\dots,\alpha_{\hat{q}_d})$ represents the eigenvectors associated with the largest $\hat{q}_d$ eigenvalues of the kernel matrix $K_{n}=(R-S)K$. The iterative algorithm is summarized in Algorithm \ref{algorithm3}.

\begin{algorithm}[htb]
	\caption{Iterative Subspace Cluster Algorithm.}
	\label{algorithm3}
	\begin{algorithmic}[1]
		\REQUIRE $ X \in \mathcal R^{n\times p}$,  $\tau=0.5$, and $c_n=0.2\log (\log(n)) \sqrt{1/n}$.
		\STATE Calculate the  $K_{0n}=RK$ and set $\hat{q}_{d0}=d-1$, then learn the basis matrix $B_{0n}$ associated with the
		$\hat{q}_{d0}$ largest eigenvalues of $K_{0n}$;
		\STATE Choose a classical clustering algorithm such as K-means to cluster the lowered data, then get $\hat{\mathcal{C}}$.
		\STATE Update the target matrix $K_n=(R-S)K$ in the formula
		(\ref{cluster}) and make the eigen-decomposition to get the eigenvalues  $\lambda_{1} \geq \ldots \geq \lambda_{n}$  and the corresponding eigenvectors ${\alpha}_1,\cdots, {\alpha}_{n}$;
		\STATE Determine the dimension $\hat{q}_{d}$ based on TRR as the formula (\ref{hatq})
		and  obtain the basis matrix $B_n=({\alpha}_1,\cdots, {\alpha}_{\hat{q}_{d}})$, then get the lowered data to be $KB_n$;
		\STATE Repeat step 2 and then calculate the RI between the clustering result and the last clustering result;
		\STATE Repeat steps 3-5 until RI exceeds 0.999.
		\ENSURE $\{\hat{\mathcal{C}}_1,\cdots,\hat{\mathcal{C}}_d\}$.\\
	\end{algorithmic}
\end{algorithm}

\section{Numerical experiments}

In this section, we perform various simulation experiments and analyze several real datasets to showcase the effectiveness of the proposed method. The numerical results pertaining to mean changes, as well as the Macroeconomic data, are provided in the Supplementary Materials in order to conserve space in the main text.

\subsection{Simulations on change point detection}

The corrected (kernel) PCA improves popular change point methods after dimension reduction. We demonstrate its effect using four popular change point detection methods: the energy-based method by \citep{matteson2014a}, the sparsified binary segmentation method \citep{cho2015multiple-change-point}, the kernel change-point algorithm \citep{kcp2019}, and the change-point detection tests using rank statistics \citep{multirank2015}, referred to as E-Divisive, SBS, KCP, and Multirank, respectively. To compare the performance of the corrected (kernel) PCA with other dimension reduction technologies, we compare three dimension reduction methods: the corrected (kernel) PCA, (kernel) PCA and the corrected Mahalanobis matrix method proposed by \cite{zhu2022}. For simplicity, we denote the versions of E-Divisive based on the dimension reduction methods as E-Divisive$_{C}$, E-Divisive$_{P}$, and E-Divisive$_{M}$, respectively. Although SBS is applicable to multivariate data,
when $\hat{q}=1$, SBS automatically reduces to wild binary segmentation method (WBS) in \cite{fryzlewicz2014wild}. We also compare the proposed method with three popular high-dimensional methods: the informative sparse projection for estimation of change points \citep{wang2018high}, the double CUSUM statistic method \citep{Cho2016Change}, and the method via a geometrically inspired mapping \citep{grundy2020high-dimensional}, referred to as Inspect, DCBS, and GeomCP, respectively. Since Inspect, DCBS, and GeomCP are applied in high-dimensional scenarios, we did not report their results after dimension reduction. We evaluate the performance of the estimated change points by measuring the average of $\hat{s}$, the root-mean-square error (RMSE) of $\hat{s}$, and the Rand index (RI) \citep{rand1971objective} between the estimated and real segments. The E-Divisive method is implemented in the R package ``ecp'', while SBS, WBS, and Inspect are implemented in the R packages ``hdbinseg'', ``wbs'', and ``InspectChangepoint'', respectively. The Python code for the Multirank method is provided by the authors of \cite{multirank2015}.

Each simulation is repeated 1000 times. The TRR method is used to select $\hat{q}_d$ for CKPCA and the Mahalanobis matrix, with parameters $\tau=0.5$ and $c_n=0.2\log (\log(n)) \sqrt{1/n}$. For CPCA, we set $c_n=0.2\log (\log(n)) \sqrt{p/n}$. The cumulative variance contribution rate method is employed to select $\hat{q}_d$ for (kernel) PCA, with a cumulative variance contribution rate of 0.95. We choose the Gaussian kernel function given by
$K(X, X')=\exp \left[-\|X-X\|^{2} /\left(2 h^{2}\right)\right],$
where $h$ represents the bandwidth, and $h>0$. Inspired by the idea of \cite{varon2015noise}, we select the bandwidth $h$ as $h^2=m \times p \times \mathbb{E}\left[\operatorname{Var}\left[X \right]\right]$ where  $\operatorname{Var}\left[X \right]=\left[\operatorname{Var}\left[X_1 \right]\right.$, $\left.\operatorname{Var}\left[X_2 \right], \ldots, \operatorname{Var}\left[X_p \right]\right]$, $\mathbb{E}\left[\operatorname{Var}\left[X \right]\right]=\frac{1}{p}\sum_{i=1}^{p} \operatorname{Var}\left[X_i \right]$, $\operatorname{Var}\left[X_i \right]$ denotes to the variance of one dimension in the dataset, and $m$ denotes a tuning parameter. We recommend $m=0.8$. More details about  the sensitivity of $m$ can be found later.

We evaluate the methods through three scenarios: (1) changes in both distribution and covariance matrix, (2) changes in distribution, and (3) changes in mean. \textbf{Examples 1-3} correspond to these scenarios, respectively. \textbf{Example 3} which pertains to changes in mean is provided in the Supplementary Materials to economize space in the main context. We apply PCA and CPCA to detect mean changes, and KPCA and CKPCA to detect distributional changes. Since SBS, WBS, Inspect, DCBS, and GeomCP are designed to detect mean changes, we only consider E-Divisive, KCP, and Multirank in \textbf{Examples 1} and \textbf{2}. However, based on the inference presented in Section 3, CKPCA can transform distributional changes into mean changes. Therefore, we also explore the performance of SBS after applying the three dimension reduction versions. The sample size is set to $n=800$. The data are divided into eight parts, each following a distribution denoted by $\{G_i\}_{i=1}^{8}$. Therefore, the total number of change points is $s=7$. We conduct experiments on both balanced and imbalanced datasets:
\begin{itemize}
	\item {\bf Balanced dataset.} The change points are located at $100i$ for $i=1,2,3,\cdots,7$, respectively;
	\item {\bf Imbalanced dataset.} The change points are located at 30, 170, 350, 440, 520, 630, 710, respectively.
\end{itemize}

\textbf{Example 1:  Changes in both distribution and covariance matrix.} The data are generated in the following settings:

\begin{itemize}
	\item Case 1: $G_1=G_3=G_5=G_7=N(0_{p}, \Sigma)$, where
	$\Sigma=(1.5\textrm{I}_{p\times p}+\sigma_{ij})$ and $\sigma_{ij}=I(i=j)+ bI(i \neq j)$ with $b=0.5$,
	and $G_2=G_4=G_6=G_8$ are the $p$-dimentional uniform distributions on the regions $[-3,3]\times [-3,3]\times \cdots\times [-3,3]$.
	\item Case 2: The settings of $G_i$ are the same as Case 1, except that $\Sigma=(1.5\textrm{I}_{p\times p}+\sigma_{ij})$ and $\sigma_{ij}=b^{\left|i-j\right|}$ with $b=0.5$.
\end{itemize}
In this example, we consider a dimension of $p=100$ and $p=200$, with a structure dimension of $q_d=1$. The results are shown in Tables \ref{Table3} and \ref{Table3u}. In Case 1, Multirank$_{C}$ performs the best, with $\hat{s}$ being close to the true value of 7 and the RI exceeding 0.99. Among the SBS methods, SBS$_{C}$ outperforms both SBS$_{P}$ and SBS$_{M}$. All three versions of dimension reduction demonstrate significant improvements for E-Divisive, with CKPCA exhibiting the most substantial enhancement. The results of Case 2 are similar to Case 1, with Multirank$_{C}$ still being the top performer. KCP and E-Divisive are less effective, but KCP$_{C}$ and E-Divisive$_{C}$ still yield good results. The RI of Multirank is lower than that of Multirank$_{C}$, and SBS$_{C}$ outperforms both SBS$_{P}$ and SBS$_{M}$.

To assess the sensitivity of the methods to outliers, we introduce imbalanced data with 5\% outliers from $G_i+W_i$ between each $z_{i}$ and $z_{i+1}$. Here, $W_i$ represents a $p$-dimensional constant vector. For each $i$, we randomly select 5\% of its elements to take the value 5, while the other elements are set to 0. The results presented in Table \ref{Table3uout} demonstrate that the dimension reduction-based methods are relatively robust against imbalanced data and data with outliers.

\begin{table}[htb!]
	\caption{Changes in both distribution and covariance matrix in \textbf{Example 1} with balanced dataset}
	\resizebox{\textwidth}{!}{
		\begin{tabular}{@{}l|lllll|lllll@{}}
			\hline
			Case & $p$ & method           & $\hat{s}$ & RMSE  & RI    & $p$ & method           & $\hat{s}$ & RMSE  & RI    \\ \hline
			\multirow{15}{*}{1}      & \multirow{15}{*}{200}   & E-Divisive$_{C}$ & 7.576     & 0.951 & 0.991 & \multirow{15}{*}{100}   & E-Divisive$_{C}$ & 7.511     & 0.904 & 0.991 \\
			&                         & E-Divisive$_{P}$ & 4.921     & 2.749 & 0.832 &                         & E-Divisive$_{P}$ & 2.590     & 4.812 & 0.555 \\
			&                         & E-Divisive$_{M}$ & 8.450     & 2.352 & 0.913 &                         & E-Divisive$_{M}$ & 6.786     & 2.070 & 0.850 \\
			&                         & E-Divisive       & 0.747     & 6.361 & 0.267 &                         & E-Divisive       & 0.693     & 6.406 & 0.268 \\
			&                         & KCP$_{C}$        & 7.140     & 1.315 & 0.977 &                         & KCP$_{C}$        & 6.057     & 2.556 & 0.879 \\
			&                         & KCP$_{P}$        & 0.421     & 6.762 & 0.181 &                         & KCP$_{P}$        & 0.000     & 7.000 & 0.124 \\
			&                         & KCP$_{M}$        & 2.653     & 5.952 & 0.396 &                         & KCP$_{M}$        & 2.394     & 5.839 & 0.408 \\
			&                         & KCP              & 0.004     & 6.996 & 0.124 &                         & KCP              & 0.030     & 6.982 & 0.128 \\
			&                         & Multirank$_{C}$  & 7.000     & 0.000 & 0.995 &                         & Multirank$_{C}$  & 7.000     & 0.000 & 0.995 \\
			&                         & Multirank$_{P}$  & 0.000     & 7.000 & 0.124 &                         & Multirank$_{P}$  & 0.000     & 7.000 & 0.124 \\
			&                         & Multirank$_{M}$  & 0.000     & 7.000 & 0.124 &                         & Multirank$_{M}$  & 0.000     & 7.000 & 0.124 \\
			&                         & Multirank        & 0.511     & 6.761 & 0.131 &                         & Multirank        & 0.336     & 6.829 & 0.132 \\
			&                         & SBS$_{C}$        & 8.359     & 2.044 & 0.994 &                         & SBS$_{C}$        & 7.939     & 1.512 & 0.994 \\
			&                         & SBS$_{P}$        & 0.039     & 6.965 & 0.135 &                         & SBS$_{P}$        & 0.039     & 6.964 & 0.134 \\
			&                         & SBS$_{M}$        & 6.392     & 3.065 & 0.772 &                         & SBS$_{M}$        & 5.044     & 3.329 & 0.729 \\ \hline
			\multirow{15}{*}{2}      & \multirow{15}{*}{200}   & E-Divisive$_{C}$ & 9.452     & 2.936 & 0.973 & \multirow{15}{*}{100}   & E-Divisive$_{C}$ & 8.824     & 2.274 & 0.972 \\
			&                         & E-Divisive$_{P}$ & 4.455     & 3.554 & 0.748 &                         & E-Divisive$_{P}$ & 0.644     & 6.450 & 0.262 \\
			&                         & E-Divisive$_{M}$ & 10.385    & 4.007 & 0.701 &                         & E-Divisive$_{M}$ & 7.783     & 2.045 & 0.869 \\
			&                         & E-Divisive       & 0.352     & 6.692 & 0.207 &                         & E-Divisive       & 0.195     & 6.828 & 0.173 \\
			&                         & KCP$_{C}$        & 8.071     & 1.752 & 0.983 &                         & KCP$_{C}$        & 6.572     & 1.605 & 0.936 \\
			&                         & KCP$_{P}$        & 0.000     & 7.000 & 0.124 &                         & KCP$_{P}$        & 0.000     & 7.000 & 0.124 \\
			&                         & KCP$_{M}$        & 11.552    & 7.342 & 0.863 &                         & KCP$_{M}$        & 5.527     & 5.558 & 0.650 \\
			&                         & KCP              & 0.000     & 7.000 & 0.124 &                         & KCP              & 0.000     & 7.000 & 0.124 \\
			&                         & Multirank$_{C}$  & 7.000     & 0.000 & 0.988 &                         & Multirank$_{C}$  & 6.987     & 0.177 & 0.982 \\
			&                         & Multirank$_{P}$  & 0.000     & 7.000 & 0.124 &                         & Multirank$_{P}$  & 0.000     & 7.000 & 0.124 \\
			&                         & Multirank$_{M}$  & 0.000     & 7.000 & 0.124 &                         & Multirank$_{M}$  & 0.000     & 7.000 & 0.124 \\
			&                         & Multirank        & 0.081     & 6.958 & 0.126 &                         & Multirank        & 0.090     & 6.948 & 0.125 \\
			&                         & SBS$_{C}$        & 10.564    & 4.235 & 0.975 &                         & SBS$_{C}$        & 8.663     & 2.396 & 0.977 \\
			&                         & SBS$_{P}$        & 0.011     & 6.990 & 0.129 &                         & SBS$_{P}$        & 0.028     & 6.974 & 0.131 \\
			&                         & SBS$_{M}$        & 9.956     & 4.050 & 0.880 &                         & SBS$_{M}$        & 6.331     & 2.226 & 0.814 \\ \hline
		\end{tabular}
		\label{Table3}
	}
\end{table}

\begin{table}[htb!]
	\caption{Changes in both distribution and covariance matrix in \textbf{Example 1} with imbalanced dataset}
	\resizebox{\textwidth}{!}{
		\begin{tabular}{@{}l|lllll|lllll@{}}
			\hline
			Case & $p$ & method           & $\hat{s}$ & RMSE  & RI    & $p$ & method           & $\hat{s}$ & RMSE  & RI    \\ \hline
			\multirow{15}{*}{1}      & \multirow{15}{*}{200}   & E-Divisive$_{C}$ & 7.636     & 1.147 & 0.988 & \multirow{15}{*}{100}   & E-Divisive$_{C}$ & 7.555     & 0.988 & 0.988 \\
			&                         & E-Divisive$_{P}$ & 3.102     & 4.185 & 0.795 &                         & E-Divisive$_{P}$ & 2.698     & 4.494 & 0.748 \\
			&                         & E-Divisive$_{M}$ & 8.549     & 2.674 & 0.897 &                         & E-Divisive$_{M}$ & 7.067     & 2.012 & 0.855 \\
			&                         & E-Divisive       & 1.007     & 6.137 & 0.346 &                         & E-Divisive       & 0.922     & 6.201 & 0.337 \\
			&                         & KCP$_{C}$        & 6.794     & 1.149 & 0.974 &                         & KCP$_{C}$        & 6.748     & 1.299 & 0.967 \\
			&                         & KCP$_{P}$        & 1.931     & 5.562 & 0.512 &                         & KCP$_{P}$        & 0.015     & 6.987 & 0.150 \\
			&                         & KCP$_{M}$        & 3.824     & 5.393 & 0.507 &                         & KCP$_{M}$        & 3.519     & 5.620 & 0.501 \\
			&                         & KCP              & 0.000     & 7.000 & 0.146 &                         & KCP              & 0.000     & 7.000 & 0.146 \\
			&                         & Multirank$_{C}$  & 6.996     & 0.067 & 0.994 &                         & Multirank$_{C}$  & 6.996     & 0.067 & 0.994 \\
			&                         & Multirank$_{P}$  & 0.000     & 7.000 & 0.146 &                         & Multirank$_{P}$  & 0.000     & 7.000 & 0.146 \\
			&                         & Multirank$_{M}$  & 0.000     & 7.000 & 0.146 &                         & Multirank$_{M}$  & 0.000     & 7.000 & 0.146 \\
			&                         & Multirank        & 0.354     & 6.821 & 0.151 &                         & Multirank        & 0.215     & 6.875 & 0.150 \\
			&                         & SBS$_{C}$        & 9.122     & 2.964 & 0.987 &                         & SBS$_{C}$        & 8.006     & 1.580 & 0.990 \\
			&                         & SBS$_{P}$        & 0.050     & 6.954 & 0.163 &                         & SBS$_{P}$        & 0.050     & 6.955 & 0.163 \\
			&                         & SBS$_{M}$        & 7.039     & 3.327 & 0.802 &                         & SBS$_{M}$        & 4.928     & 3.343 & 0.721 \\ \hline
			\multirow{15}{*}{2}      & \multirow{15}{*}{200}   & E-Divisive$_{C}$ & 9.779     & 3.210 & 0.961 & \multirow{15}{*}{100}   & E-Divisive$_{C}$ & 9.128     & 2.667 & 0.961 \\
			&                         & E-Divisive$_{P}$ & 4.004     & 3.509 & 0.832 &                         & E-Divisive$_{P}$ & 0.672     & 6.406 & 0.319 \\
			&                         & E-Divisive$_{M}$ & 10.436    & 4.058 & 0.884 &                         & E-Divisive$_{M}$ & 7.672     & 2.250 & 0.852 \\
			&                         & E-Divisive       & 0.399     & 6.649 & 0.249 &                         & E-Divisive       & 0.161     & 6.855 & 0.193 \\
			&                         & KCP$_{C}$        & 6.794     & 1.438 & 0.968 &                         & KCP$_{C}$        & 6.405     & 2.027 & 0.922 \\
			&                         & KCP$_{P}$        & 0.000     & 7.000 & 0.146 &                         & KCP$_{P}$        & 0.000     & 7.000 & 0.146 \\
			&                         & KCP$_{M}$        & 8.076     & 4.539 & 0.779 &                         & KCP$_{M}$        & 4.908     & 4.910 & 0.628 \\
			&                         & KCP              & 0.000     & 7.000 & 0.146 &                         & KCP              & 0.000     & 7.000 & 0.146 \\
			&                         & Multirank$_{C}$  & 6.901     & 0.367 & 0.983 &                         & Multirank$_{C}$  & 6.686     & 0.715 & 0.970 \\
			&                         & Multirank$_{P}$  & 0.000     & 7.000 & 0.146 &                         & Multirank$_{P}$  & 0.000     & 7.000 & 0.146 \\
			&                         & Multirank$_{M}$  & 0.000     & 7.000 & 0.146 &                         & Multirank$_{M}$  & 0.000     & 7.000 & 0.146 \\
			&                         & Multirank        & 0.117     & 6.943 & 0.150 &                         & Multirank        & 0.090     & 6.958 & 0.150 \\
			&                         & SBS$_{C}$        & 11.685    & 5.236 & 0.960 &                         & SBS$_{C}$        & 9.287     & 3.030 & 0.963 \\
			&                         & SBS$_{P}$        & 0.044     & 6.959 & 0.164 &                         & SBS$_{P}$        & 0.011     & 6.990 & 0.148 \\
			&                         & SBS$_{M}$        & 9.702     & 3.890 & 0.867 &                         & SBS$_{M}$        & 6.144     & 2.561 & 0.800 \\ \hline
		\end{tabular}
		\label{Table3u}
	}
\end{table}

\begin{table}[htb!]
	\caption{Changes in both distribution and covariance matrix in \textbf{Example 1} with outliers}
	\resizebox{\textwidth}{!}{
		\begin{tabular}{@{}l|lllll|lllll@{}}
			\hline
			Case & $p$ & method           & $\hat{s}$ & RMSE  & RI    & $p$ & method           & $\hat{s}$ & RMSE  & RI    \\ \hline
			\multirow{15}{*}{1}      & \multirow{15}{*}{200}   & E-Divisive$_{C}$ & 7.655     & 1.211 & 0.988 & \multirow{15}{*}{100}   & E-Divisive$_{C}$ & 7.602     & 1.074 & 0.987 \\
			&                         & E-Divisive$_{P}$ & 3.069     & 4.154 & 0.793 &                         & E-Divisive$_{P}$ & 2.490     & 4.651 & 0.732 \\
			&                         & E-Divisive$_{M}$ & 7.931     & 2.287 & 0.886 &                         & E-Divisive$_{M}$ & 6.847     & 2.157 & 0.845 \\
			&                         & E-Divisive       & 1.054     & 6.082 & 0.362 &                         & E-Divisive       & 0.805     & 6.297 & 0.315 \\
			&                         & KCP$_{C}$        & 6.870     & 0.833 & 0.984 &                         & KCP$_{C}$        & 6.389     & 1.969 & 0.933 \\
			&                         & KCP$_{P}$        & 1.695     & 5.776 & 0.459 &                         & KCP$_{P}$        & 0.000     & 7.000 & 0.146 \\
			&                         & KCP$_{M}$        & 2.511     & 5.797 & 0.416 &                         & KCP$_{M}$        & 2.221     & 5.601 & 0.430 \\
			&                         & KCP              & 0.000     & 7.000 & 0.146 &                         & KCP              & 0.000     & 7.000 & 0.146 \\
			&                         & Multirank$_{C}$  & 7.000     & 0.000 & 0.994 &                         & Multirank$_{C}$  & 7.000     & 0.000 & 0.994 \\
			&                         & Multirank$_{P}$  & 0.000     & 7.000 & 0.124 &                         & Multirank$_{P}$  & 0.000     & 7.000 & 0.124 \\
			&                         & Multirank$_{M}$  & 0.000     & 7.000 & 0.124 &                         & Multirank$_{M}$  & 0.000     & 7.000 & 0.124 \\
			&                         & Multirank        & 0.677     & 6.647 & 0.140 &                         & Multirank        & 0.215     & 6.887 & 0.131 \\
			&                         & SBS$_{C}$        & 8.464     & 2.176 & 0.989 &                         & SBS$_{C}$        & 7.950     & 1.688 & 0.991 \\
			&                         & SBS$_{P}$        & 0.028     & 6.974 & 0.158 &                         & SBS$_{P}$        & 0.061     & 6.944 & 0.165 \\
			&                         & SBS$_{M}$        & 6.884     & 3.161 & 0.814 &                         & SBS$_{M}$        & 5.177     & 3.102 & 0.757 \\ \hline
			\multirow{15}{*}{2}      & \multirow{15}{*}{200}   & E-Divisive$_{C}$ & 9.785     & 3.250 & 0.959 & \multirow{15}{*}{100}   & E-Divisive$_{C}$ & 9.234     & 2.700 & 0.956 \\
			&                         & E-Divisive$_{P}$ & 5.398     & 2.173 & 0.925 &                         & E-Divisive$_{P}$ & 1.747     & 5.499 & 0.527 \\
			&                         & E-Divisive$_{M}$ & 10.084    & 3.799 & 0.882 &                         & E-Divisive$_{M}$ & 7.525     & 2.127 & 0.857 \\
			&                         & E-Divisive       & 1.349     & 5.780 & 0.482 &                         & E-Divisive       & 0.625     & 6.455 & 0.308 \\
			&                         & KCP$_{C}$        & 7.084     & 0.961 & 0.977 &                         & KCP$_{C}$        & 8.237     & 2.168 & 0.968 \\
			&                         & KCP$_{P}$        & 0.000     & 7.000 & 0.146 &                         & KCP$_{P}$        & 0.000     & 7.000 & 0.146 \\
			&                         & KCP$_{M}$        & 5.847     & 3.829 & 0.724 &                         & KCP$_{M}$        & 4.756     & 4.498 & 0.668 \\
			&                         & KCP              & 0.000     & 7.000 & 0.146 &                         & KCP              & 0.000     & 7.000 & 0.146 \\
			&                         & Multirank$_{C}$  & 7.000     & 0.000 & 0.986 &                         & Multirank$_{C}$  & 6.960     & 0.307 & 0.979 \\
			&                         & Multirank$_{P}$  & 0.000     & 7.000 & 0.124 &                         & Multirank$_{P}$  & 0.000     & 7.000 & 0.124 \\
			&                         & Multirank$_{M}$  & 0.000     & 7.000 & 0.124 &                         & Multirank$_{M}$  & 0.000     & 7.000 & 0.124 \\
			&                         & Multirank        & 0.045     & 6.967 & 0.126 &                         & Multirank        & 0.135     & 6.924 & 0.128 \\
			&                         & SBS$_{C}$        & 10.414    & 4.078 & 0.965 &                         & SBS$_{C}$        & 8.989     & 2.711 & 0.965 \\
			&                         & SBS$_{P}$        & 0.028     & 6.974 & 0.155 &                         & SBS$_{P}$        & 0.039     & 6.964 & 0.159 \\
			&                         & SBS$_{M}$        & 9.006     & 3.495 & 0.850 &                         & SBS$_{M}$        & 6.022     & 2.591 & 0.797 \\ \hline
		\end{tabular}
		\label{Table3uout}
	}
\end{table}

\textbf{Example 2:  Changes in distribution.} The data are generated in the following settings:
\begin{itemize}
	\item $G_1=G_3=G_5=G_7=N(0_{p}, \Sigma)$ and $G_2=G_4=G_6=G_8=t(df, \Sigma)$ is the $p$-dimentional t-distribution with the degree $df=a$ and $\Sigma=(\sigma_{ij})$, where  $\sigma_{ij}=0.5^{\left|i-j\right|}$, $a=4, 6$, $p=100, 200$.
\end{itemize}

In \textbf{Example 2}, the distributions change while the mean and covariance remain constant. We consider different values of $a=4$ and $a=6$, representing weak and strong signals, respectively. The results of \textbf{Example 2} are presented in Tables \ref{Table2} and \ref{Table2u}. When $a=6$, E-Divisive and KCP are almost ineffective, while E-Divisive$_{C}$ and KCP$_{C}$ still perform well. Additionally, Multirank$_{C}$ outperforms Multirank, and SBS$_{C}$ outperforms both SBS$_{P}$ and SBS$_{M}$. The results when $a=4$ are similar to those when $a=6$. All three methods show improvement after applying CKPCA. \textbf{Example 3} can be found in the Supplementary Materials for the purpose of conserving space. Overall, \textbf{Examples 1-3} demonstrate that CPCA/CKPCA can enhance the performance of popular change point detection methods, surpassing PCA/KPCA. Furthermore, based on the results from imbalanced cases, outliers, and distinct distributions, CPCA/CKPCA exhibits greater robustness compared to its competitors.

\begin{table}[htb!]
	\caption{Changes in distribution in \textbf{Example 2} with balanced dataset}
	\resizebox{\textwidth}{!}{
		\begin{tabular}{@{}l|lllll|llllll@{}}
			\hline
			$p$ & $a$ & method           & $\hat{s}$ & RMSE   & RI    & $p$ & $a$ & method           & $\hat{s}$ & RMSE   & RI    \\ \hline
			\multirow{15}{*}{200}   & \multirow{15}{*}{4}     & E-Divisive$_{C}$ & 7.293       & 0.639  & 0.974 & \multirow{15}{*}{100}   & \multirow{15}{*}{4}     & E-Divisive$_{C}$ & 7.176       & 0.822  & 0.964 \\
			&                         & E-Divisive$_{P}$ & 6.461       & 1.154  & 0.936 &                         &                         & E-Divisive$_{P}$ & 6.030       & 1.655  & 0.903 \\
			&                         & E-Divisive$_{M}$ & 2.744       & 4.949  & 0.515 &                         &                         & E-Divisive$_{M}$ & 2.448       & 5.016  & 0.509 \\
			&                         & E-Divisive       & 4.255       & 3.353  & 0.745 &                         &                         & E-Divisive       & 3.324       & 4.166  & 0.652 \\
			&                         & KCP$_{C}$        & 10.061      & 7.352  & 0.940 &                         &                         & KCP$_{C}$        & 8.454       & 6.451  & 0.857 \\
			&                         & KCP$_{P}$        & 4.748       & 4.480  & 0.682 &                         &                         & KCP$_{P}$        & 3.950       & 5.052  & 0.578 \\
			&                         & KCP$_{M}$        & 1.316       & 7.681  & 0.188 &                         &                         & KCP$_{M}$        & 5.504       & 10.148 & 0.331 \\
			&                         & KCP              & 0.450       & 7.345  & 0.138 &                         &                         & KCP              & 1.054       & 7.778  & 0.159 \\
			&                         & Multirank$_{C}$  & 4.309       & 4.127  & 0.635 &                         &                         & Multirank$_{C}$  & 3.605       & 4.690  & 0.560 \\
			&                         & Multirank$_{P}$  & 0.099       & 6.970  & 0.124 &                         &                         & Multirank$_{P}$  & 1.117       & 6.656  & 0.143 \\
			&                         & Multirank$_{M}$  & 0.000       & 7.000  & 0.124 &                         &                         & Multirank$_{M}$  & 0.000       & 7.000  & 0.124 \\
			&                         & Multirank        & 1.327       & 6.428  & 0.161 &                         &                         & Multirank        & 0.682       & 6.725  & 0.143 \\
			&                         & SBS$_{C}$        & 8.646       & 3.469  & 0.891 &                         &                         & SBS$_{C}$        & 7.923       & 2.899  & 0.852 \\
			&                         & SBS$_{P}$        & 0.017       & 6.985  & 0.127 &                         &                         & SBS$_{P}$        & 0.022       & 6.979  & 0.131 \\
			&                         & SBS$_{M}$        & 9.619       & 5.688  & 0.677 &                         &                         & SBS$_{M}$        & 8.293       & 3.909  & 0.627 \\ \hline
			\multirow{15}{*}{200}   & \multirow{15}{*}{6}     & E-Divisive$_{C}$ & 7.433       & 1.140  & 0.961 & \multirow{15}{*}{100}   & \multirow{15}{*}{6}     & E-Divisive$_{C}$ & 6.324       & 2.481  & 0.863 \\
			&                         & E-Divisive$_{P}$ & 4.931       & 2.665  & 0.814 &                         &                         & E-Divisive$_{P}$ & 4.251       & 3.390  & 0.747 \\
			&                         & E-Divisive$_{M}$ & 4.608       & 3.715  & 0.679 &                         &                         & E-Divisive$_{M}$ & 3.471       & 4.146  & 0.630 \\
			&                         & E-Divisive       & 1.981       & 5.290  & 0.479 &                         &                         & E-Divisive       & 0.784       & 6.326  & 0.284 \\
			&                         & KCP$_{C}$        & 11.218      & 10.356 & 0.737 &                         &                         & KCP$_{C}$        & 9.500       & 9.932  & 0.640 \\
			&                         & KCP$_{P}$        & 1.224       & 6.394  & 0.276 &                         &                         & KCP$_{P}$        & 0.031       & 6.972  & 0.129 \\
			&                         & KCP$_{M}$        & 2.794       & 8.007  & 0.282 &                         &                         & KCP$_{M}$        & 5.084       & 9.285  & 0.371 \\
			&                         & KCP              & 0.851       & 7.639  & 0.150 &                         &                         & KCP              & 6.850       & 11.151 & 0.339 \\
			&                         & Multirank$_{C}$  & 3.910       & 4.413  & 0.597 &                         &                         & Multirank$_{C}$  & 3.215       & 5.016  & 0.508 \\
			&                         & Multirank$_{P}$  & 0.000       & 7.000  & 0.124 &                         &                         & Multirank$_{P}$  & 0.000       & 7.000  & 0.124 \\
			&                         & Multirank$_{M}$  & 0.000       & 7.000  & 0.124 &                         &                         & Multirank$_{M}$  & 0.000       & 7.000  & 0.124 \\
			&                         & Multirank        & 0.937       & 6.616  & 0.144 &                         &                         & Multirank        & 0.426       & 6.831  & 0.134 \\
			&                         & SBS$_{C}$        & 7.845       & 3.271  & 0.731 &                         &                         & SBS$_{C}$        & 6.481       & 3.056  & 0.644 \\
			&                         & SBS$_{P}$        & 0.017       & 6.985  & 0.129 &                         &                         & SBS$_{P}$        & 0.028       & 6.975  & 0.132 \\
			&                         & SBS$_{M}$        & 7.779       & 3.958  & 0.663 &                         &                         & SBS$_{M}$        & 5.718       & 3.260  & 0.587 \\ \hline
		\end{tabular}
		\label{Table2}	}
\end{table}

\begin{table}[htb!]
	\caption{Changes in distribution in \textbf{Example 2} with imbalanced dataset}
	\resizebox{\textwidth}{!}{
		\begin{tabular}{@{}l|lllll|llllll@{}}
			\hline
			$p$ & $a$ & method           & $\hat{s}$ & RMSE   & RI    & $p$ & $a$ & method           & $\hat{s}$ & RMSE  & RI    \\ \hline
			\multirow{15}{*}{200}   & \multirow{15}{*}{4}     & E-Divisive$_{C}$ & 7.165       & 0.685  & 0.973 & \multirow{15}{*}{100}   & \multirow{15}{*}{4}     & E-Divisive$_{C}$ & 6.924       & 1.019 & 0.959 \\
			&                         & E-Divisive$_{P}$ & 5.603       & 1.879  & 0.921 &                         &                         & E-Divisive$_{P}$ & 5.174       & 2.283 & 0.893 \\
			&                         & E-Divisive$_{M}$ & 2.889       & 4.904  & 0.521 &                         &                         & E-Divisive$_{M}$ & 2.380       & 5.099 & 0.509 \\
			&                         & E-Divisive       & 3.501       & 3.845  & 0.749 &                         &                         & E-Divisive       & 2.720       & 4.643 & 0.632 \\
			&                         & KCP$_{C}$        & 10.321      & 7.551  & 0.947 &                         &                         & KCP$_{C}$        & 9.321       & 8.048 & 0.822 \\
			&                         & KCP$_{P}$        & 4.489       & 4.315  & 0.710 &                         &                         & KCP$_{P}$        & 0.053       & 6.957 & 0.155 \\
			&                         & KCP$_{M}$        & 1.198       & 7.633  & 0.201 &                         &                         & KCP$_{M}$        & 0.947       & 7.285 & 0.198 \\
			&                         & KCP              & 0.000       & 7.000  & 0.146 &                         &                         & KCP              & 0.000       & 7.000 & 0.146 \\
			&                         & Multirank$_{C}$  & 3.587       & 4.553  & 0.594 &                         &                         & Multirank$_{C}$  & 3.462       & 4.671 & 0.579 \\
			&                         & Multirank$_{P}$  & 0.143       & 6.941  & 0.149 &                         &                         & Multirank$_{P}$  & 0.807       & 6.614 & 0.158 \\
			&                         & Multirank$_{M}$  & 0.000       & 7.000  & 0.146 &                         &                         & Multirank$_{M}$  & 0.000       & 7.000 & 0.146 \\
			&                         & Multirank        & 1.439       & 6.425  & 0.170 &                         &                         & Multirank        & 0.771       & 6.663 & 0.167 \\
			&                         & SBS$_{C}$        & 8.392       & 2.973  & 0.887 &                         &                         & SBS$_{C}$        & 7.017       & 2.925 & 0.796 \\
			&                         & SBS$_{P}$        & 0.011       & 6.990  & 0.148 &                         &                         & SBS$_{P}$        & 0.022       & 6.979 & 0.152 \\
			&                         & SBS$_{M}$        & 10.083      & 5.811  & 0.659 &                         &                         & SBS$_{M}$        & 8.431       & 4.216 & 0.643 \\ \hline
			\multirow{15}{*}{200}   & \multirow{15}{*}{6}     & E-Divisive$_{C}$ & 7.076       & 1.224  & 0.960 & \multirow{15}{*}{100}   & \multirow{15}{*}{6}     & E-Divisive$_{C}$ & 5.774       & 2.873 & 0.843 \\
			&                         & E-Divisive$_{P}$ & 4.223       & 3.153  & 0.821 &                         &                         & E-Divisive$_{P}$ & 3.484       & 3.913 & 0.735 \\
			&                         & E-Divisive$_{M}$ & 4.653       & 3.653  & 0.686 &                         &                         & E-Divisive$_{M}$ & 3.362       & 4.217 & 0.619 \\
			&                         & E-Divisive       & 1.438       & 5.771  & 0.427 &                         &                         & E-Divisive       & 0.620       & 6.461 & 0.282 \\
			&                         & KCP$_{C}$        & 12.076      & 11.466 & 0.714 &                         &                         & KCP$_{C}$        & 6.290       & 8.367 & 0.534 \\
			&                         & KCP$_{P}$        & 1.481       & 6.507  & 0.307 &                         &                         & KCP$_{P}$        & 0.000       & 7.000 & 0.146 \\
			&                         & KCP$_{M}$        & 3.557       & 8.345  & 0.329 &                         &                         & KCP$_{M}$        & 2.618       & 8.476 & 0.263 \\
			&                         & KCP              & 0.000       & 7.000  & 0.146 &                         &                         & KCP              & 0.000       & 7.000 & 0.146 \\
			&                         & Multirank$_{C}$  & 3.265       & 4.813  & 0.544 &                         &                         & Multirank$_{C}$  & 2.578       & 5.373 & 0.456 \\
			&                         & Multirank$_{P}$  & 0.000       & 7.000  & 0.146 &                         &                         & Multirank$_{P}$  & 0.000       & 7.000 & 0.146 \\
			&                         & Multirank$_{M}$  & 0.000       & 7.000  & 0.146 &                         &                         & Multirank$_{M}$  & 0.000       & 7.000 & 0.146 \\
			&                         & Multirank        & 0.534       & 6.765  & 0.160 &                         &                         & Multirank        & 0.422       & 6.833 & 0.159 \\
			&                         & SBS$_{C}$        & 7.414       & 3.584  & 0.705 &                         &                         & SBS$_{C}$        & 5.740       & 3.247 & 0.609 \\
			&                         & SBS$_{P}$        & 0.017       & 6.985  & 0.153 &                         &                         & SBS$_{P}$        & 0.006       & 6.995 & 0.148 \\
			&                         & SBS$_{M}$        & 8.238       & 4.343  & 0.662 &                         &                         & SBS$_{M}$        & 6.061       & 3.077 & 0.602 \\ \hline
		\end{tabular}
		\label{Table2u}	}
\end{table}

We test the method's sensitivity to bandwidth selection by considering values of  $m=0.4, 0.8, 1.2, 1.6, 2.0$ for \textbf{Example 2}. Table \ref{Table7} presents the E-Divisive results for different bandwidth values when $a=4$ and $p=200$. The results demonstrate robustness across bandwidth variations. In Table \ref{Table7}, the best $\hat{s}$ results occur with $m=1.6$, while the RMSE and RI perform optimally at $m=0.8$. Therefore, selecting $m=0.8$ in simulation studies is reasonable.

\begin{table}[htb!]
	\begin{center}
		\caption{The results of \textbf{Example 2} with different bandwidth values}
		\label{Table7}
		\begin{tabular}{ccllll}
			\hline
			$p$ & $a$ & $m$ & $\hat{s}$ & RMSE  & RI    \\ \hline
			\multirow{5}{*}{200}    & \multirow{5}{*}{4}      & 0.4 & 7.566     & 1.076 & 0.974 \\
			&                         & 0.8 & 7.257     & 0.548 & 0.975 	\\
			&                         & 1.2 & 7.199     & 0.612 & 0.971 	\\
			&                         & 1.6 & 7.145     & 0.701 & 0.968 	\\
			&                         & 2.0   & 6.897     & 1.011 & 0.952		\\ \hline
		\end{tabular}
	\end{center}
\end{table}

\begin{figure}[htb!]
	\centering
	\includegraphics[width=5.25cm,height=4.5cm]{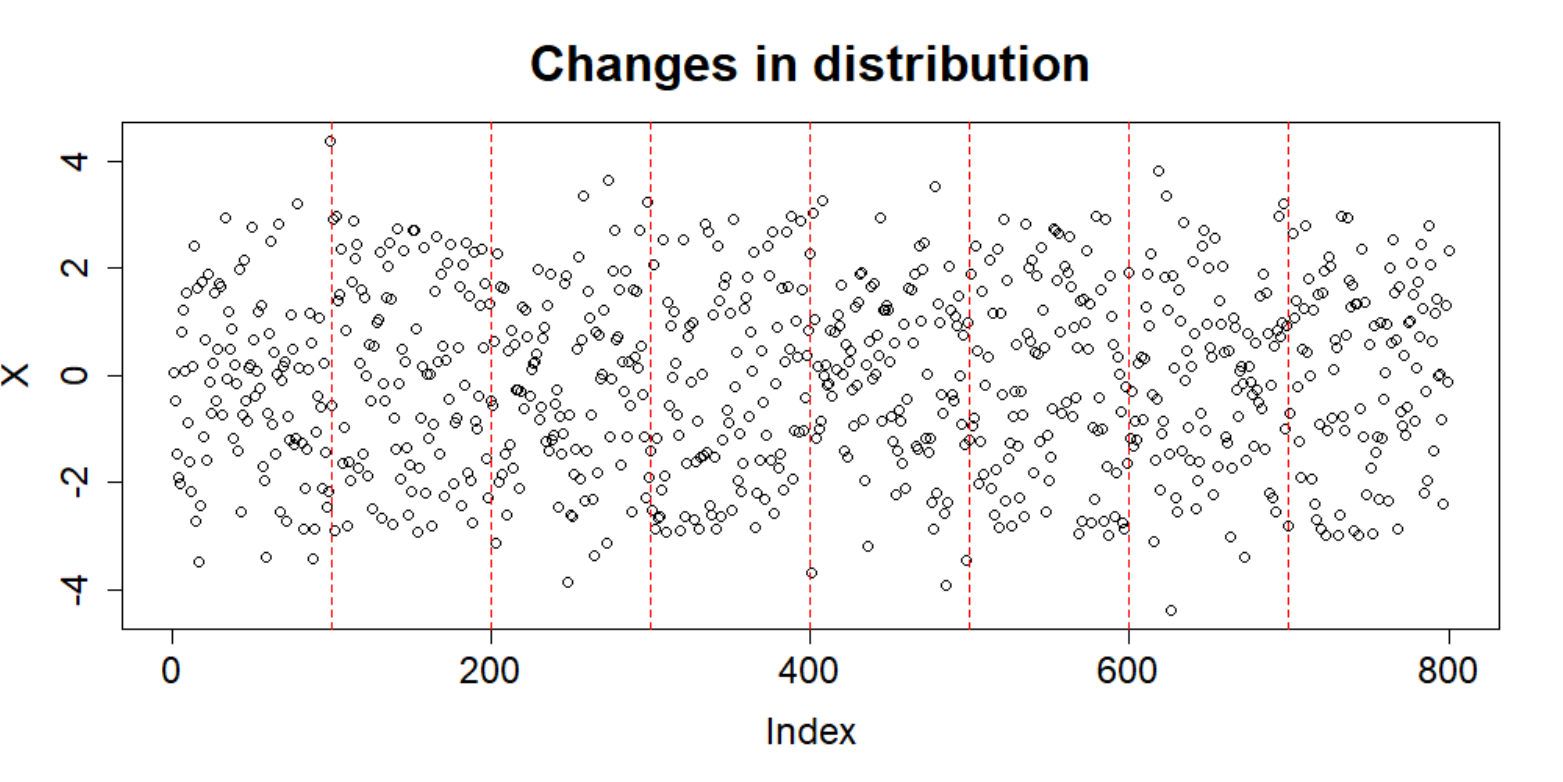}
	\centering
	\includegraphics[width=5.25cm,height=4.5cm]{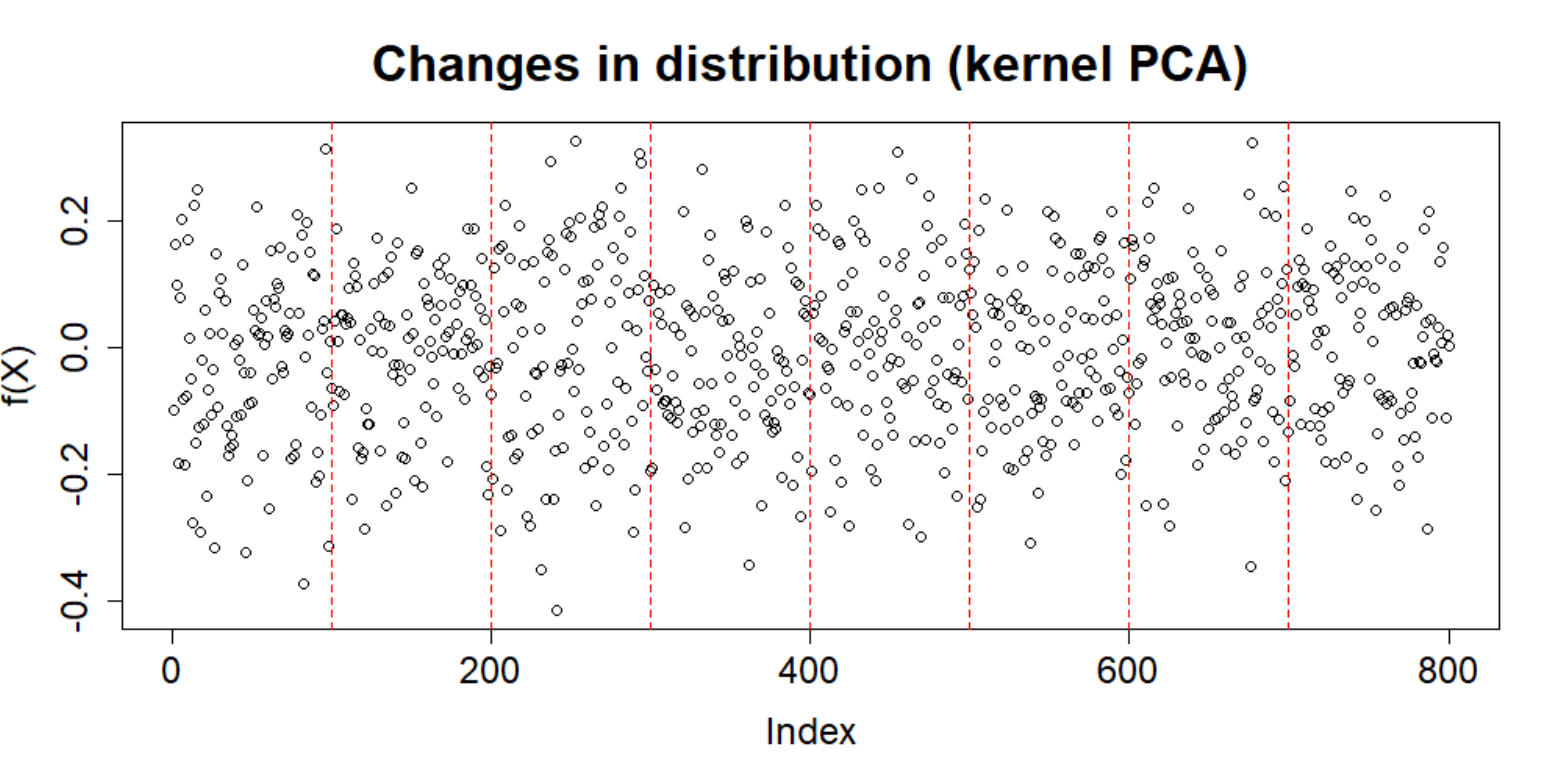}
	\centering
	\includegraphics[width=5.25cm,height=4.5cm]{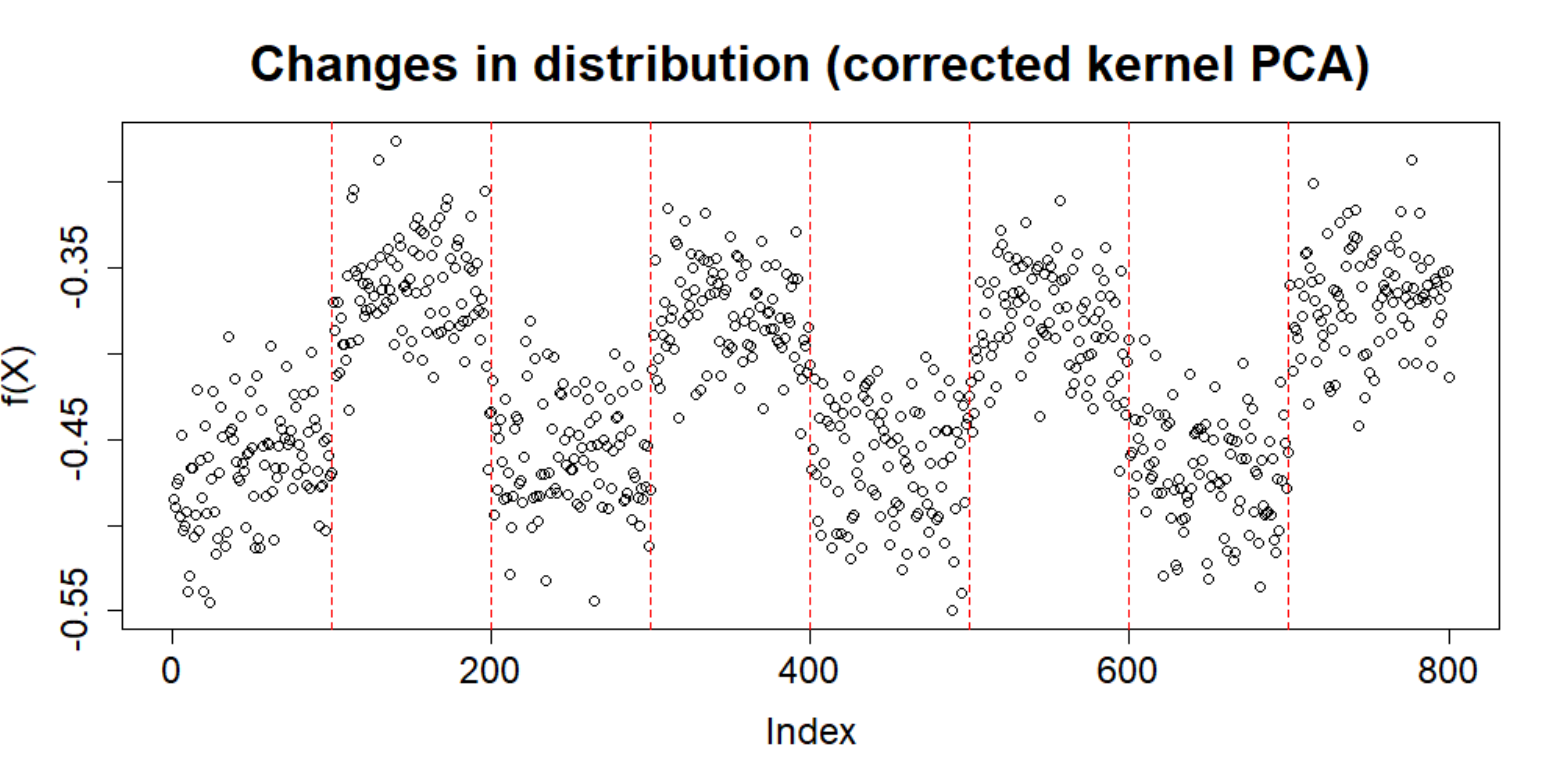}
	\vspace{-0.5cm}
	\caption{
Scatter plots present the first vectors of the original data and the data after KPCA and CKPCA, respectively. The three figures correspond to Case 2 in \textbf{Example 1} with $p=200$.
	}
	\label{fig1}
\end{figure}

To gain intuitive understanding of CKPCA, we plot scatter plots in Figure \ref{fig1} for the first vector of the original data, $\{B^{\top}_{P1n}X_i\}_{i=1}^n$, and $\{B^{\top}_{C1n}X_i\}_{i=1}^n$. Here, $B^{\top}_{C1n}$ and $B^{\top}_{P1n}$ represent the first vector of the data after CKPCA and KPCA, respectively. Figure \ref{fig1} clearly shows that the changes at the change points become more pronounced after CKPCA, while KPCA does not facilitate change point detection.

\subsection{Simulations on clustering}

We conducted a comparative analysis between the iterative subspace cluster algorithm and several well-known clustering methods applied directly on datasets that exhibit clustering structure. Specifically, we considered four commonly used clustering methods: the K-means method \citep{kmeans}, the partitioning around medoid method \citep{PAM}, the expectation-maximization algorithm \citep{EM}, and the density-based spatial clustering of applications with noise method \citep{dbscan}, which are denoted as K-means, PAM, EM, and DBSCAN, respectively. We compared two versions of dimension reduction techniques: Corrected Kernel PCA (CKPCA) and Kernel PCA (KPCA). For example, K-means$_{C}$ and K-means$_{P}$ represent the iterative CKPCA method and KPCA based on K-means, respectively. To assess the performance, we employed the Rand Index (RI) as a measure of similarity between the underlying clusters and the estimated clusters. The average (mean) and standard deviation (sd) of the RI values were reported. The bandwidth $h$ is selected following the procedure described for change point detection.
Experiments are conducted on balanced and imbalanced datasets with three categories: (1) balanced dataset with equal sample sizes $n_1=n_2=n_3=n/3$; (2) imbalanced dataset with sample sizes $n_1=300$, $n_2=200$, and $n_3=100$. Data $X$ is generated according to the following settings: the $k$th class contains $\{X_{k,i}\}_{i=1}^{n_k}$, where $X_{k,i}=\sigma_{k,i}w_{k,i}$ for $k=1,2,3$ and $i=1,2,\cdots,n_{k}$, with $\sigma_{k,i}$ sampled from a uniform distribution in the range $[2k-2,2k-1]$, and $w_{k,i}$ sampled from a uniform distribution on the unit sphere $\mathbb{S}^{p}$. Here, $p=10, 100$. Each simulation is repeated 1000 times.

The findings are presented in Tables \ref{Table5} and \ref{Table5u}. Among the various methods considered for balanced data, it is observed that four CKPCA versions exhibit superior performance, achieving a Rand Index (RI) of 1. However, the effectiveness of DBSCAN is limited in this particular example, whereas DBSCAN$_{C}$ demonstrates favorable performance. Notably, CKPCA significantly enhances the performance of most clustering methods, with iterative CKPCA surpassing KPCA. Similar patterns emerge when examining the results for imbalanced data, with CKPCA versions displaying the most favorable performance. Conversely, DBSCAN performs poorly, but its performance is improved by DBSCAN$_{C}$. These results unequivocally indicate that iterative CKPCA enhances the performance of popular clustering methods and demonstrates exceptional robustness in the context of imbalanced cases.

\begin{table}[htb!]
	\begin{center}
		\caption{Balanced data with $n_1=n_2=n_3=200$}
		\begin{tabular}{c|lllc|lll}
			\hline
			$p$ & method        & mean  & sd    & $p$ & method        & mean  & sd    \\ \hline
			\multirow{12}{*}{100}   & K-means$_{C}$ & 1.000 & 0.000 & \multirow{12}{*}{10}    & K-means$_{C}$ & 1.000 & 0.000 \\
			& K-means       & 0.505 & 0.015 &                         & K-means       & 0.563 & 0.013 \\
			& K-means$_{P}$ & 1.000 & 0.000 &                         & K-means$_{P}$ & 0.999 & 0.020 \\
			& PAM$_{C}$     & 1.000 & 0.000 &                         & PAM$_{C}$     & 1.000 & 0.000 \\
			& PAM           & 0.334 & 0.000 &                         & PAM           & 0.523 & 0.027 \\
			& PAM$_{P}$     & 1.000 & 0.000 &                         & PAM$_{P}$     & 0.922 & 0.014 \\
			& EM$_{C}$      & 1.000 & 0.000 &                         & EM$_{C}$      & 1.000 & 0.000 \\
			& EM            & 0.733 & 0.007 &                         & EM            & 0.734 & 0.006 \\
			& EM$_{P}$      & 0.739 & 0.007 &                         & EM$_{P}$      & 0.731 & 0.008 \\
			& DBSCAN$_{C}$  & 1.000 & 0.000 &                         & DBSCAN$_{C}$  & 1.000 & 0.000 \\
			& DBSCAN        & 0.387 & 0.009 &                         & DBSCAN        & 0.401 & 0.012 \\
			& DBSCAN$_{P}$  & 0.484 & 0.016 &                         & DBSCAN$_{P}$  & 0.384 & 0.010 \\ \hline
		\end{tabular}
		\label{Table5}
	\end{center}
\end{table}

\begin{table}[htb!]
	\begin{center}
		\caption{Imbalanced data with $n_1=300$, $n_2=200$, $n_3=100$}
		\begin{tabular}{c|lllc|lll}
			\hline
			$p$ & method        & mean  & sd    & $p$ & method        & mean  & sd    \\ \hline
			\multirow{12}{*}{100}   & K-means$_{C}$ & 0.963 & 0.084 & \multirow{12}{*}{10}    & K-means$_{C}$ & 0.985 & 0.059 \\
			& K-means       & 0.540 & 0.020 &                         & K-means       & 0.623 & 0.016 \\
			& K-means$_{P}$ & 0.928 & 0.106 &                         & K-means$_{P}$ & 0.982 & 0.057 \\
			& PAM$_{C}$     & 1.000 & 0.000 &                         & PAM$_{C}$     & 1.000 & 0.000 \\
			& PAM           & 0.392 & 0.000 &                         & PAM           & 0.639 & 0.036 \\
			& PAM$_{P}$     & 1.000 & 0.000 &                         & PAM$_{P}$     & 0.993 & 0.005 \\
			& EM$_{C}$      & 1.000 & 0.000 &                         & EM$_{C}$      & 1.000 & 0.000 \\
			& EM            & 0.780 & 0.008 &                         & EM            & 0.794 & 0.010 \\
			& EM$_{P}$      & 0.791 & 0.010 &                         & EM$_{P}$      & 0.783 & 0.012 \\
			& DBSCAN$_{C}$  & 1.000 & 0.000 &                         & DBSCAN$_{C}$  & 1.000 & 0.000 \\
			& DBSCAN        & 0.409 & 0.005 &                         & DBSCAN        & 0.423 & 0.008 \\
			& DBSCAN$_{P}$  & 0.467 & 0.012 &                         & DBSCAN$_{P}$  & 0.404 & 0.005 \\ \hline
		\end{tabular}
		\label{Table5u}
	\end{center}
\end{table}

\subsection{Real data}

\subsubsection{Genetics data with mean changes}

%
%

We analyze an array comparative genomic hybridization (aCGH) microarray dataset, previously analyzed in \cite{Stransky2006Regional} and \cite{Blaveri2006Bladder}, to detect mean changes in the data structure. The dataset comprises 57 individuals with bladder tumors. We utilize the processed data from the R package ``ecp'' and select 43 individuals out of the 57, along with 2215 different loci on their genome, resulting in $p=43$ and $n=2215$. In the bandwidth formula, we set $m=0.8$. This empirical study aims to identify unusual chromosomal characteristics.

\begin{figure}[htb!]
	\centering
	\includegraphics[width=15cm,height=4cm]{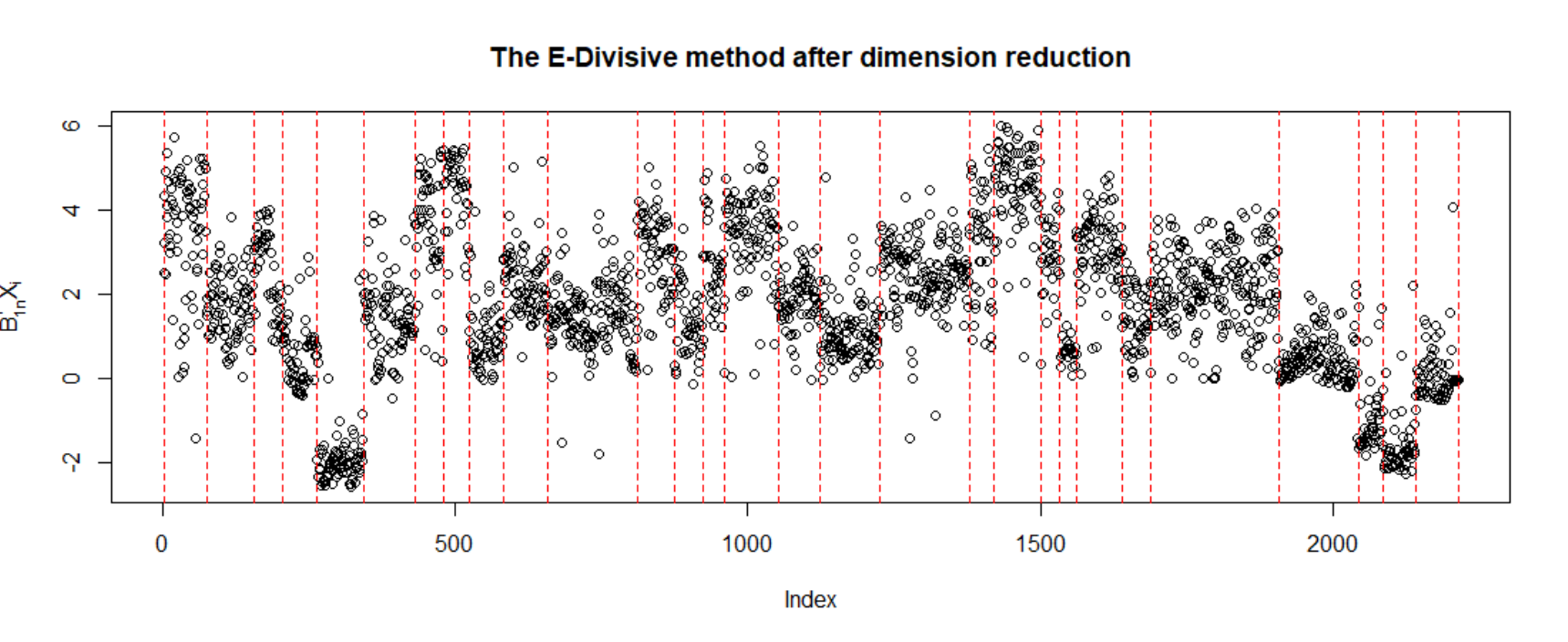}
	\vspace{-0.5cm}
	\caption{Change point detection for aCGH data, the figure plots the locations detected by the dimension reduction-based E-divisive method.}
	\label{real_data3}
\end{figure}

Given that E-Divisive$_{C}$ outperformed other methods in previous simulation studies, we employ this method. We discover 28 change points in the dimension-reduced data. Since the dimension $q_d$ is determined as $1$ based on the TRR criterion, Figure \ref{real_data3} depicts the change point locations.	From Figure \ref{real_data3}, it is evident that we can successfully detect the jump locations in the data.

\subsubsection{The clustering data}

\begin{figure}[htb!]
	\centering
	\includegraphics[width=3cm,height=3cm]{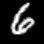}
	\quad
	\centering
	\includegraphics[width=3cm,height=3cm]{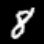}
	\quad
	\centering
	\includegraphics[width=3cm,height=3cm]{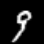}
	\vspace{-0.5cm}
	\caption{Images of handwritten digits representing the numbers 6, 8, and 9}
	\label{handwritten}
\end{figure}
\begin{figure}[htb!]
	\centering
	\includegraphics[width=4.5cm,height=4.5cm]{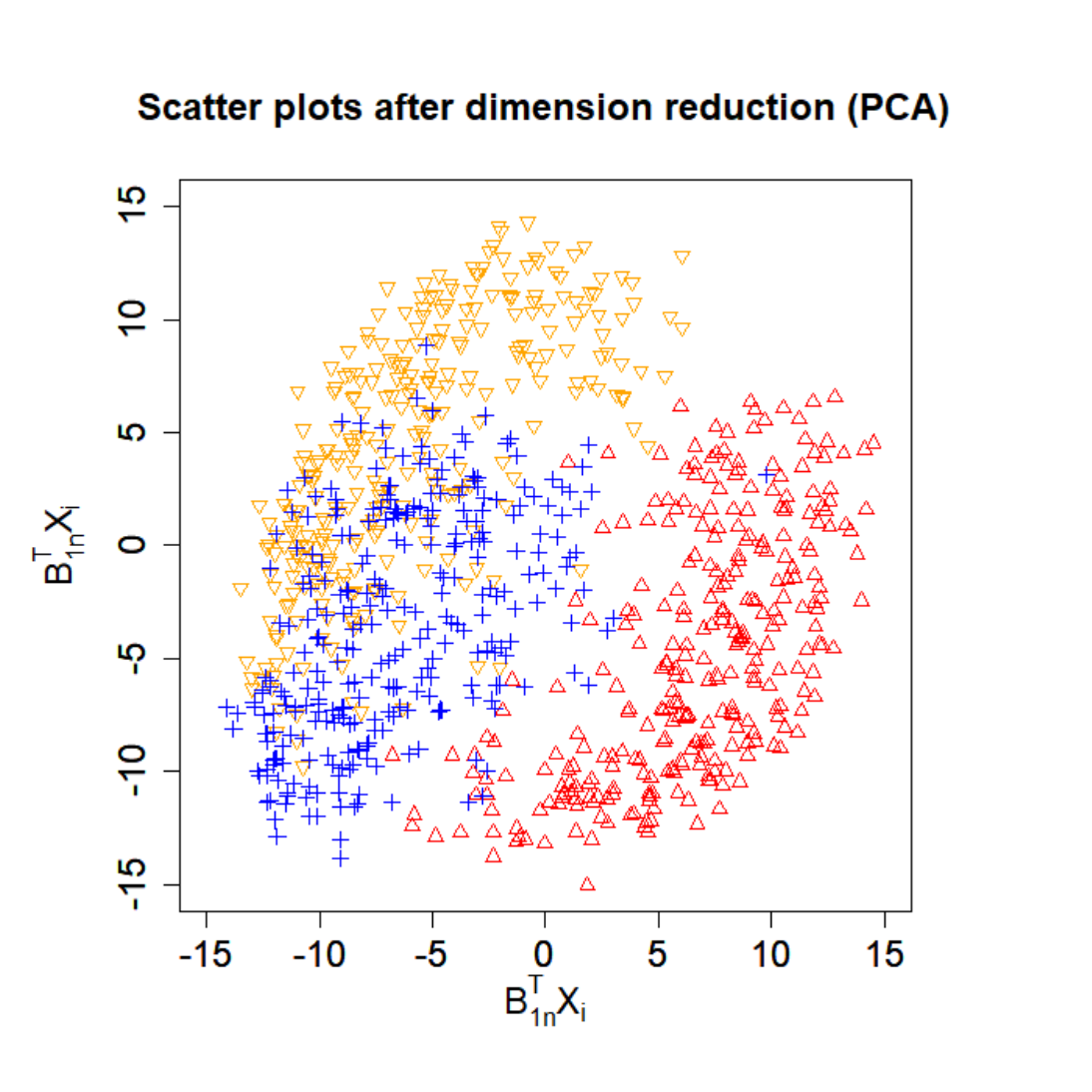}
	\centering
	\includegraphics[width=4.5cm,height=4.5cm]{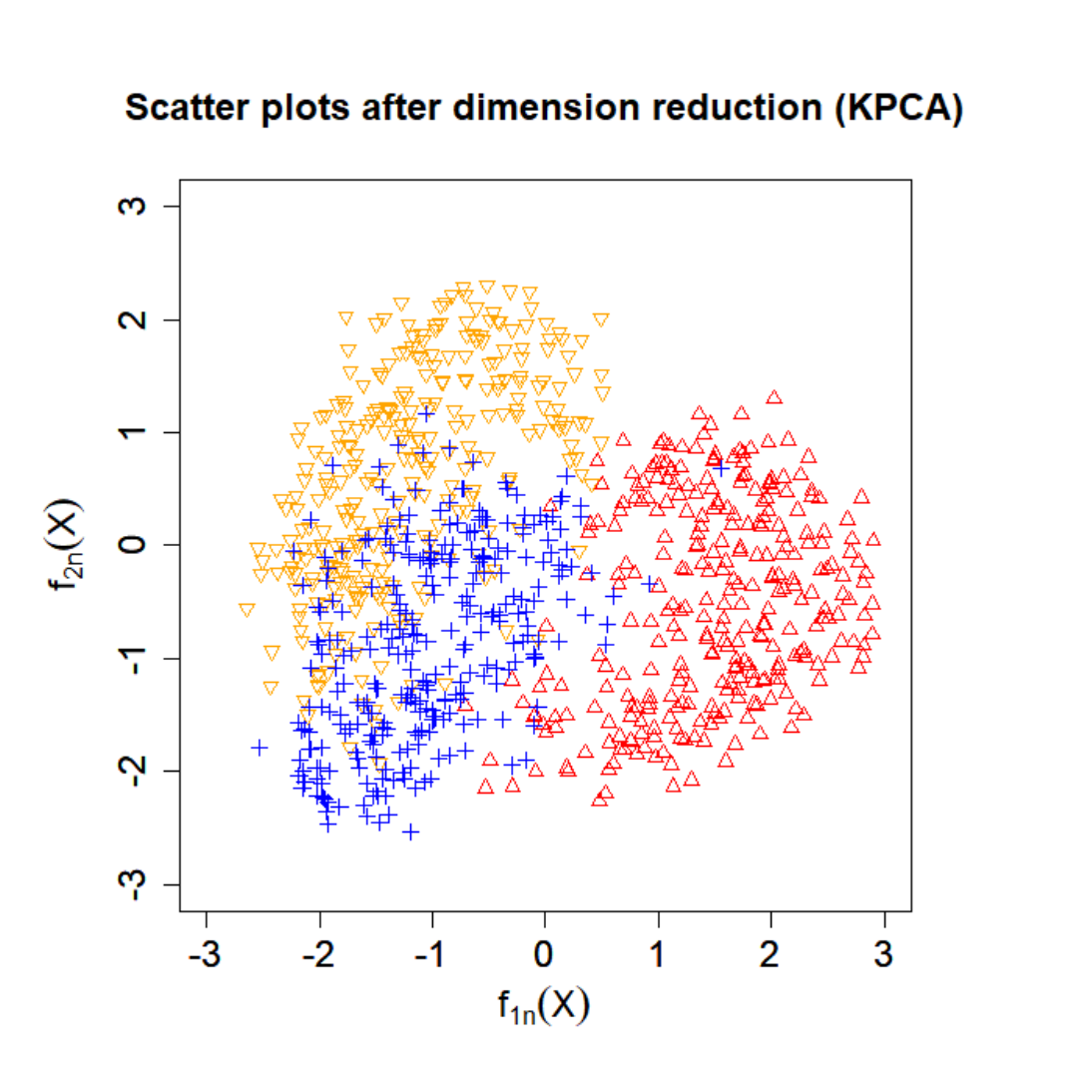}
	\centering
	\includegraphics[width=4.5cm,height=4.5cm]{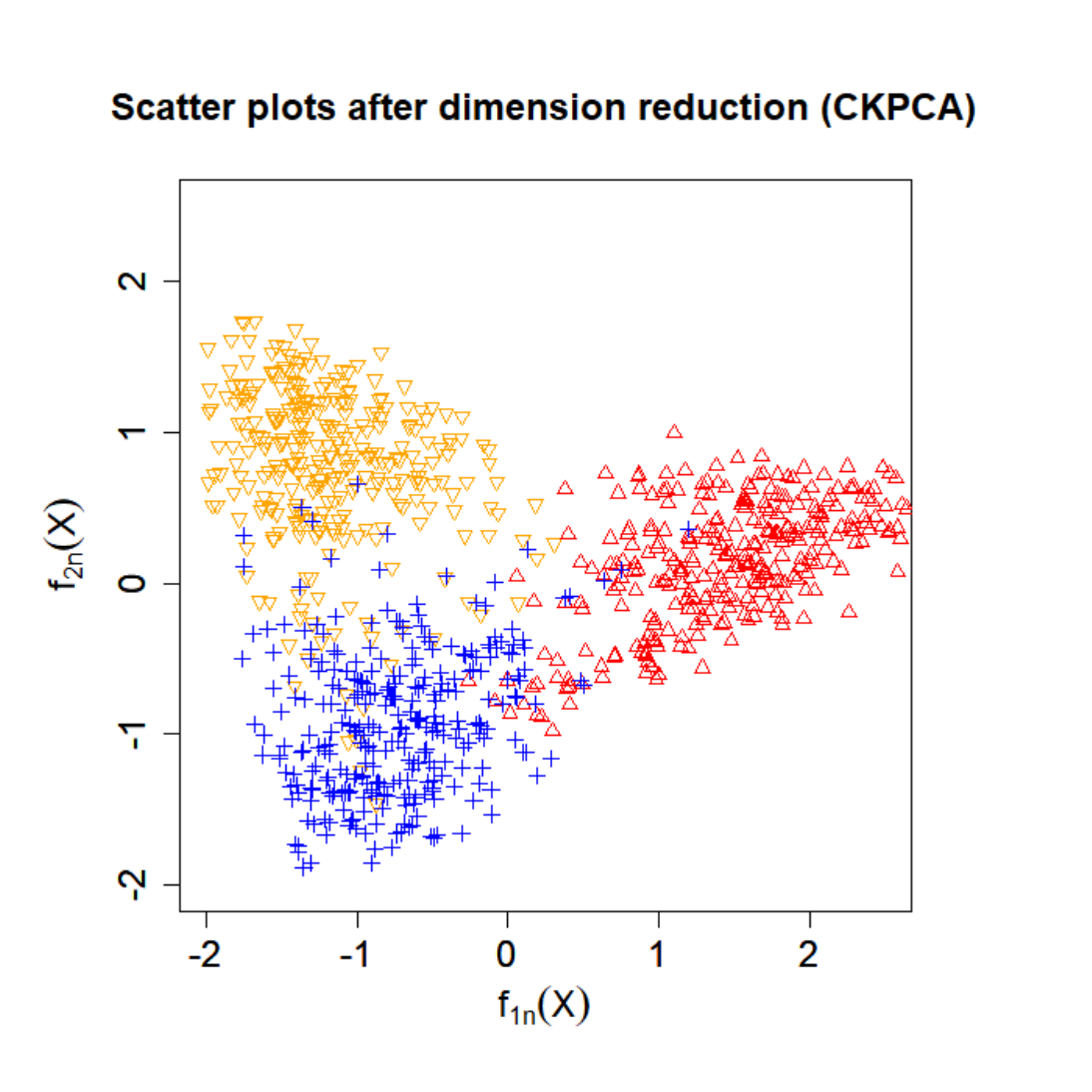}
	\vspace{-0.5cm}	
	\caption{Scatter plots after PCA, KPCA and CKPCA, the three groups of points in red, blue, and orange corresponding to the numbers 6, 8, and 9, respectively }
	\label{dimension}
\end{figure}

We utilize the MNIST (Modified National Institute of Standards and Technology) dataset, available at \url{http://yann.lecun.com/exdb/mnist/}, for clustering analysis. Specifically, we employ the proposed iterative KPCA algorithm and compare its performance against the original KPCA version. The evaluation metric used is the Rand index, which measures the similarity between the real and estimated clustering results. Our analysis focuses on a subset of the MNIST dataset consisting of 900 samples, divided into three groups of 300 samples each. Each group corresponds to handwritten digits 6, 8, and 9, which are known to be more challenging to distinguish. Figure \ref{handwritten} displays examples of these handwritten digits. Each digit is represented as a grayscale image with dimensions $28\times28$. Therefore, we have $n=900$ instances, $p=784$ dimensions, and $d=3$ categories, with $n_1=n_2=n_3=300$. To mitigate sampling randomness, we repeat the experiment 50 times.
In our experiments, we set $m=0.8$ in the bandwidth formula. Figure \ref{dimension} shows the scatter plots of the first two dimensions after PCA, KPCA, and CKPCA. It is observed that applying CKPCA improves the separability of different classes compared to KPCA. The clustering results are presented in Table \ref{Table6}. According to the results, K-means$_{C}$ and PAM$_{C}$ achieve the best performance. Additionally, applying the proposed iterative KPCA algorithm improves the performance of all four clustering methods.

\begin{table}[htb!]
	\begin{center}
		\caption{The clustering results of real data}
		\begin{tabular}{llllllll}
			\hline
			\multicolumn{8}{c}{Mnist Data}                                    \\ \hline
			method       & RI    & method   & RI    & method        & RI    & method    & RI    \\ \hline
			DBSCAN$_{C}$ & 0.704 & EM$_{C}$ & 0.737 & K-means$_{C}$ & 0.857 & PAM$_{C}$ & 0.843 \\
			DBSCAN       & 0.333 & EM       & 0.334 & K-means       & 0.809 & PAM       & 0.787 \\
			DBSCAN$_{P}$ & 0.333 & EM$_{P}$ & 0.334 & K-means$_{P}$ & 0.824 & PAM$_{P}$ & 0.789 \\ \hline
		\end{tabular}
		\label{Table6}
	\end{center}
\end{table}

\section{Conclustion}

This paper  suggests a Corrected Kernel Principal Components Analysis (CKPCA) method  and introduces a notion of central distribution deviation subspace for identifying distributional changes in high-dimensional data. The identification is implemented in this dimension reduction subspace  without loss of information on the original data.  As a special case, the corrected principal components analysis is developed to identify the central mean deviation subspace  and then  detect high-dimensional mean changes. Finally, we extend CKPCA to cluster analysis, aiming to achieve a superior nonlinear low-dimensional embedding, and produce an iterative subspace cluster algorithm.

This general  methodology can be readily extendable to other change point detection problems of high-dimensional data such as missing data \citep{follain2021high}, tensor data \citep{huang2022two}, private data \citep{Berrett2021}, online data \citep{chen2021inference}. The asymptotic results apply to both dense and sparse data structures.  The main limitation of the current method is its incapability of  handling ultra-high dimension data. A possible solution is combining a simultaneous variable selection, see, e.g., \citep{wang2018estimating,lin2019sparse,qian2019sparse}. The research is ongoing.

%
%
%

\bibliographystyle{Chicago}

\bibliography{pca_change}

\end{document}